\documentclass{aa}

\usepackage[varg]{txfonts}
\usepackage{graphicx}
\usepackage{array, booktabs, makecell}
\usepackage{siunitx}
\usepackage[version=4]{mhchem}
\usepackage{textcomp, gensymb}
\usepackage[export]{adjustbox}
\usepackage{amsmath}
\usepackage{amsfonts}
\usepackage{comment}
\usepackage{multicol}
\usepackage[utf8]{inputenc}
\usepackage[english]{babel}
\usepackage{natbib}
\bibpunct{(}{)}{;}{a}{}{,}
\usepackage[hidelinks,colorlinks=true,linkcolor=blue,citecolor=blue]{hyperref}
\usepackage[dvipsnames]{xcolor}

\usepackage{orcidlink}

\begin{document}

\title{Spectral and magnetic properties of the jet base in NGC\,315} 

\author{L.\ Ricci \orcidlink{0000-0002-4175-3194}\inst{1,2}, 
        B.\ Boccardi \orcidlink{0000-0002-4555-6390}\inst{2},
        J.\ R{\"o}der \orcidlink{0000-0002-2426-927X}\inst{2},
        M.\ Perucho \orcidlink{0000-0003-2784-0379}\inst{3,4},
        G.\ Mattia \orcidlink{0000-0003-1454-6226}\inst{5,6},
        M.\ Kadler \orcidlink{0000-0001-5606-6154}\inst{1},
        P.\ Benke \orcidlink{0009-0006-4186-9978}\inst{2},
        V.\ Bartolini \orcidlink{0009-0008-4659-2917}\inst{2},
        T.\ P.\ Krichbaum \orcidlink{0000-0002-4892-9586}\inst{2}, and
        E.\ Madika\inst{2}
     }

\institute{
\inst{1} Julius-Maximilians-Universit{\"a}t W{\"u}rzburg, Fakult{\"a}t für Physik und 
Astronomie, Institut für Theoretische Physik und Astrophysik, 
Lehrstuhl für Astronomie, Emil-Fischer-Str. 31, D-97074 W{\"u}rzburg, 
Germany \\
\inst{2} Max-Planck-Institut f{\"u}r Radioastronomie, Auf dem H{\"u}gel 69, D-53121 Bonn, Germany \\
\inst{3} Departament d’Astronomia i Astrofísica, Universitat de València, C/ Dr. Moliner, 50, 46100, Burjassot, València, Spain \\
\inst{4} Observatori Astronòmic, Universitat de València, C/ Catedràtic José Beltrán 2, 46980, Paterna, València, Spain \\
\inst{5} Max-Planck Institute for Astronomy (MPIA), K{\"o}nigstuhl 17, 69117 Heidelberg, Germany \\ 
\inst{6} INFN - Sezione di Firenze, Via G. Sansone 1, I-50019 Sesto Fiorentino (FI), Italy \\ 
}

\date{Received / Accepted}

  \abstract
   {}
{The dynamic of relativistic jets in the inner parsec regions is deeply affected by the nature of the magnetic fields. The level of magnetization of the plasma, as well as the geometry of these fields on compact scales, have not yet been fully constrained.}
{In this paper we employ multi-frequency and multi-epoch very long baseline interferometry observations of the nearby radio galaxy NGC~315. We aim to derive insights into the magnetic field properties on sub-parsec and parsec scales by examining observational signatures such as the spectral index, synchrotron turnover frequency, and brightness temperature profiles. This analysis is performed by considering the properties of the jet acceleration and collimation zone, which can be probed thanks to the source vicinity, as well as the inner part of the jet conical region.}
{
We observe remarkably steep values for the spectral index on sub-parsec scales ($\alpha \sim -2$, $S_\nu \propto \nu^\alpha$) which flatten around $\alpha \sim -0.8$ on parsec scales. 
We suggest that the observed steep values may result from particles being accelerated via diffusive shock acceleration mechanisms in magnetized plasma and subsequently experiencing cooling through synchrotron losses. 
The brightness temperature of the 43\,GHz cores indicates a dominance of the magnetic energy at the jet base, while the cores at progressively lower frequencies reveal a gradual transition towards equipartition. Based on the spectral index and brightness temperature along the incoming jet, and by employing theoretical models, we derive that the magnetic field strength has a close-to-linear dependence with distance going from parsec scales up to the jet apex.
Overall, our findings are consistent with a toroidal-dominated magnetic field on all the analyzed scales.}
   {}
   
\titlerunning{Spectral and magnetic properties of the inner parsec scales jet in NGC~315}
\authorrunning{L. Ricci et al.}
\maketitle

\section{Introduction} \label{sec:introduction}

Relativistic jets are plasma outflows launched from the direct vicinity of black holes, for instance, in active galactic nuclei (AGN).
The most energetic jets have the unique ability to develop structures that propagate up to thousands of parsecs, reaching ultra-relativistic velocities and opening angles of a few degrees \citep{Blandford2019}.
The launching of jets is a consequence of the interplay between the supermassive black hole and the magnetic fields present in the surrounding accretion disk,  
with the jets fueled by the extraction of rotational energy from the central object \citep{Blandford1977}. 
The current paradigm asserts that jets initially propagate as Poynting flux-dominated outflows; that is, their dynamics is dominated by the magnetic fields conveyed from the central region. 
The magnetic fields are expected to play a crucial role in jet propagation, especially in the so-called acceleration and collimation region.
This region extends up to $10^3 - 10^7 R_\mathrm{s}$\footnote{$R_\mathrm{S}$ is the Schwarzschild radius defined as $R_\mathrm{S} = 2 G M_\mathrm{BH} / c^2$ where $M_\mathrm{BH}$ is the mass of the black hole, $G$ is the gravitational constant, and $c$ is the speed of light.}
from the jet injection point \citep{Vlahakis2004, Marscher2008, Boccardi2021, Kovalev2020} and, within it, the jets reach small opening angles and high Lorentz factors.
This process is mainly driven by the conversion of magnetic energy into kinetic energy of the bulk flow \citep[][]{Vlahakis2003_a, Vlahakis2003_b, Vlahakis2004, Ricci2024}, in which the hoop stress generated by the toroidal field indirectly contributes to the acceleration, since it provides the recollimation and the differential expansion required by the magnetic acceleration.
Since this process was shown to become inefficient in a conical jet \citep{Komissarov2007, Komissarov2012}, the acceleration and collimation mechanisms are expected to be co-spatial.
As the magnetic field strength decreases with distance to the central engine, the acceleration and collimation process eventually stops, which is signaled by the occurrence of a transition in the jet shape from (quasi-)parabolic (namely, 
$r \propto \mathrm{z}^{\sim 0.5}$, where $\mathrm{z}$ is the radial distance from the core and $r$ is the jet radius) geometrical shape to conical $(r \propto \mathrm{z}^{\sim 1})$ \citep[the occurence of the transition region was firsly discovered by][]{Asada2012}.

Therefore, to explore the jet launching and propagation phenomena, observational constraints on the magnetic field properties on sufficiently small scales, are necessary. 
These can be probed in selected nearby targets
thanks to very long baseline interferometry (VLBI) observations performed at centimeter and millimeter wavelengths.
In such studies, magnetic fields of hundreds of mG have been inferred on parsec scales \citep[see, for example,][]{OSullivan2009, Baczko2016}, implying strengths of $10^2 - 10^4 \, \mathrm{G}$ at the jet forming region assuming the magnetic field strength to decrease with distance as $\mathrm{z}^{-1}$.
Whether this is a valid assumption is currently unknown, and exponents flatter than $\mathrm{z}^{-1}$ have been recently proposed \citep[for instance, for M\,87,][]{Ro2023}.

In this paper, we expand the magnetic field studies presented by \citet{Ricci2022} for the nearby giant radio galaxy NGC~315 \citep[$z = 0.0165$,][]{Trager2000}. 
NGC~315 is a Fanaroff--Riley Type I \citep{FaranoffRiley} morphology radio galaxy \citep[][]{Laing2006} extending $\sim 1 \, \mathrm{Mpc}$ which hosts at its center a supermassive black hole with a mass of $M_\mathrm{BH} = (2.08 \pm 0.01)^{+0.32}_{-0.14} \times 10^9 M_\odot$ \citep{Boizelle2021}. 
This target was chosen due to the abundance of available VLBI observations, and because its acceleration and collimation region has been extensively studied and its properties constrained \cite{Boccardi2021, Park2021, Ricci2022}.
We aim to further constrain the magnetic field strength and morphology on sub-parsec and parsec scales by examining 
the spatial evolution of the spectral index and of the brightness temperature.
Throughout the work, we assume a viewing angle for the source of $\theta = 38\degree$, according to \citet{Giovannini2001, Boccardi2021, Ricci2022}.




The paper is structured as follows. In Sect.~\ref{sec:observations} we describe the new data alongside the literature ones used in this work; in Sect.~\ref{sec:results} we present our results on the putative position of the central black hole, brightness temperature, and spatial evolution of the synchrotron spectrum; in Sect.~\ref{sec:discussion} we discuss the results and apply theoretical models to constraints the nature of the jet magnetic fields, and in Sect.~\ref{sec:Conclusions} we highlight our conclusions.

In this manuscript, we assume a $\Lambda$CDM cosmology $H_0 = 71 \ \mathrm{h \ km \ s^{-1} \ Mpc^{-1}}$, $\Omega_M = 0.27$, $\Omega_\Lambda = 0.73$ \citep{Komatsu}.
The luminosity distance of NGC~315 is $D_L = 70.6 \, \mathrm{Mpc}$, and 1 mas corresponds to 0.331 pc.

\begin{figure}[htpb]
    \centering
    \includegraphics[width=0.925\linewidth]{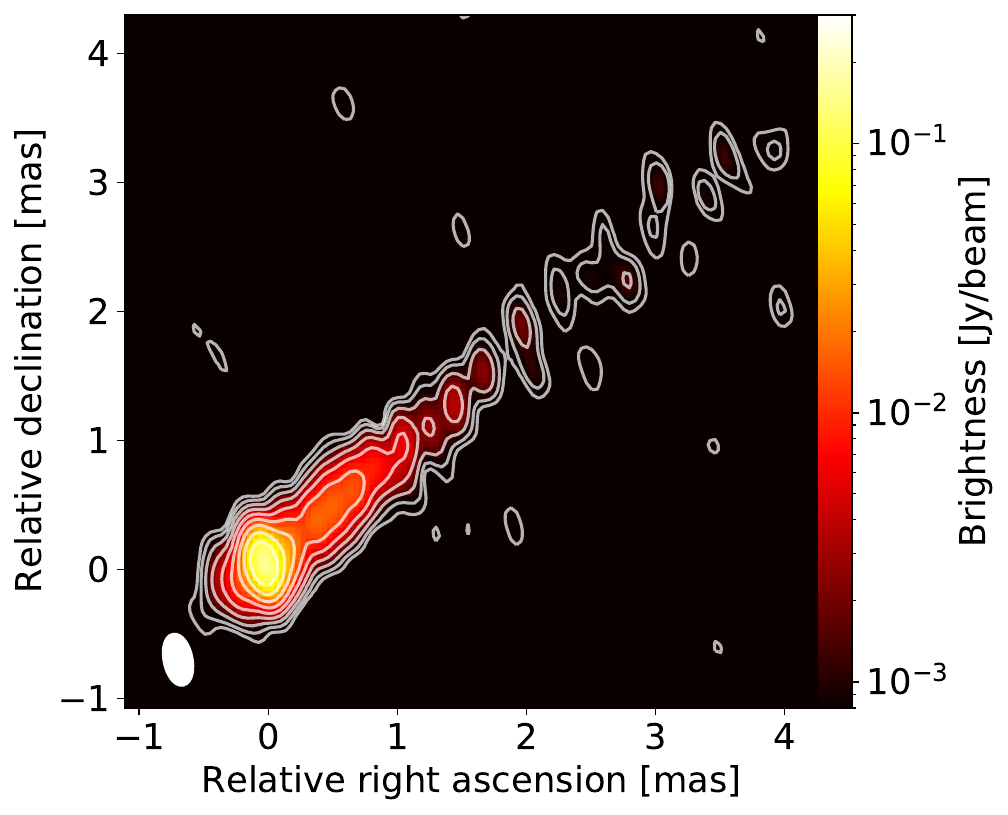}\par
    \includegraphics[width=0.925\linewidth]{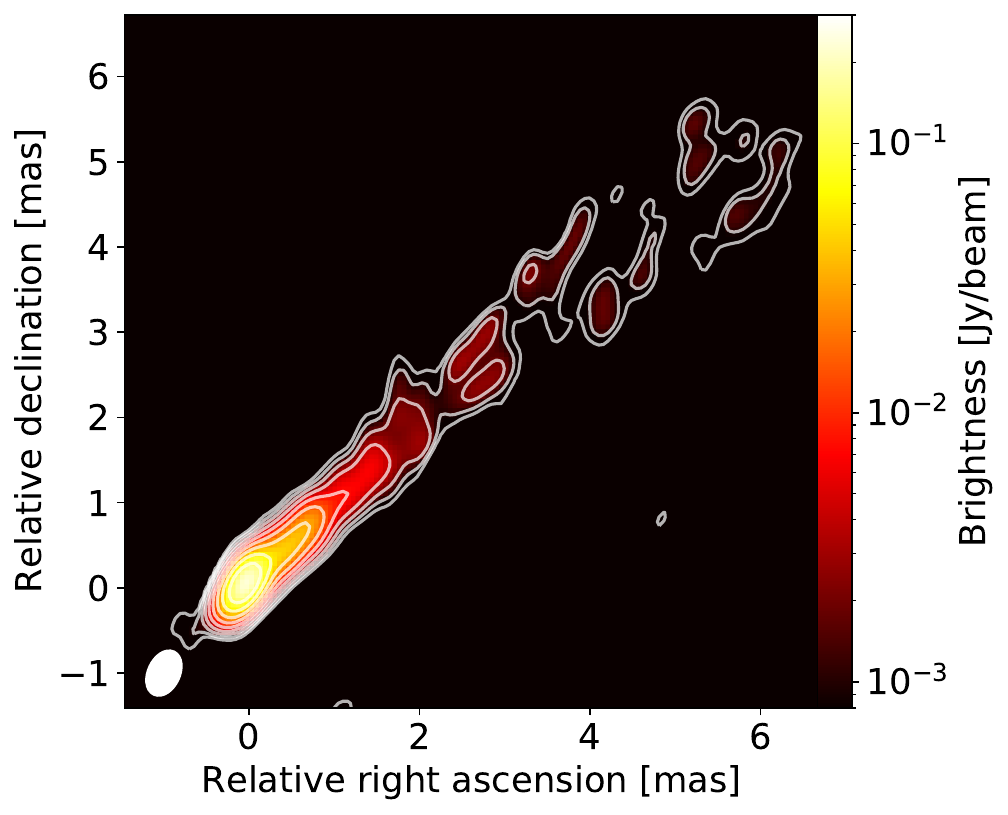}\par
    \includegraphics[width=0.925\linewidth]{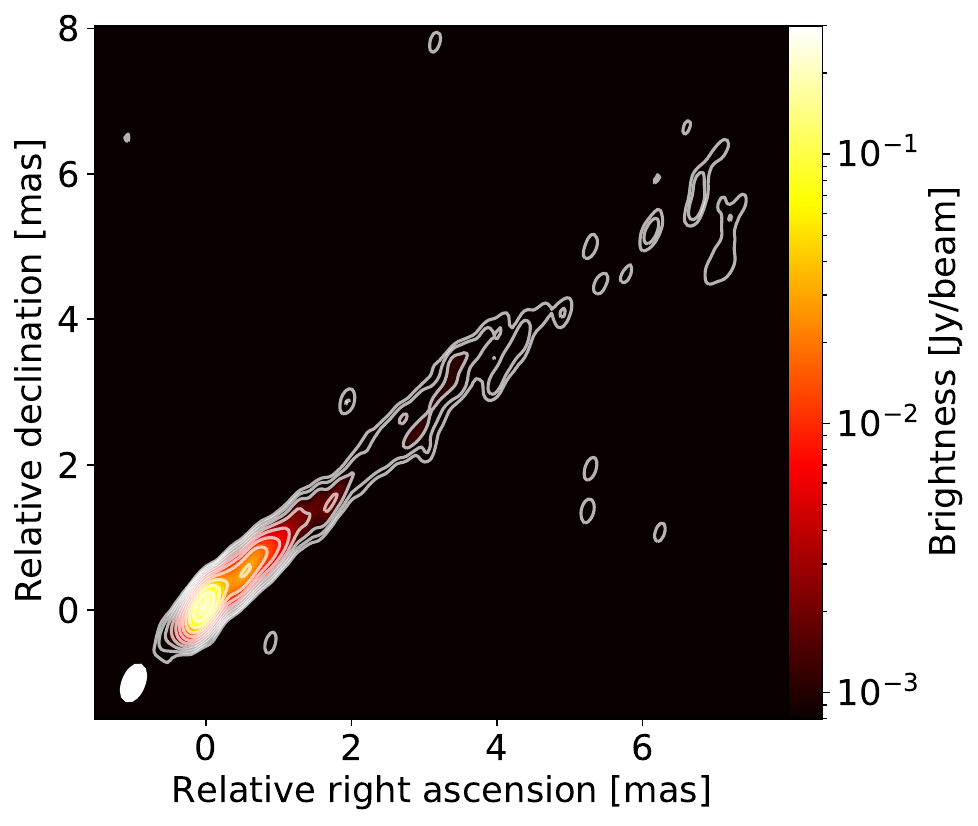}\par  
    \caption{43\,GHz VLBA observations performed in: upper panel, April 2021; middle panel, October 2021; lower panel, Apr 2022. The contours are set to (-0.2, 0.2, 0.4, 0.8, 1.6, 3.2, 6.4, 12.8, 25.6, 51.2)\% of the respective brightness peaks in each map. The images were produced with natural weighting and no taper. In all the maps, a counter-jet is detected, as expected from previous VLBI observations \citep{Boccardi2021, Park2021}. The observations performed in Oct 2021 hint at the presence of limb-brightening on $\sim 4 \, \mathrm{mas}$ scales, as suggested by \citep{Park2021} and confirmed by \citet{Park2024}.}
    \label{fig:VLBA_43GHz}
\end{figure}

\section{Data set and maps analysis} \label{sec:observations}


\subsection{Data sets} \label{sec:original_data}

\begin{table*}[htpb]
\caption{Summary of the VLBA data. The upper three lines report the properties of the new 43\,GHz data while the bottom lines include the simultaneous VLBA data set.}
\centering
\begin{tabular}{lllllll}
\hline
\multicolumn{1}{|c|}{\begin{tabular}[c]{@{}c@{}}$\nu$\\ {[}GHz{]}\end{tabular}} & 
\multicolumn{1}{|c|}{Project code} & 
\multicolumn{1}{|c|}{Obs date} & 
\multicolumn{1}{|c|}{\begin{tabular}[c]{@{}c@{}}Beam\\ {[}mas, mas, deg{]}\end{tabular}} & \multicolumn{1}{c|}{\begin{tabular}[c]{@{}c@{}}Total flux\\ {[}Jy{]}\end{tabular}} & \multicolumn{1}{c|}{\begin{tabular}[c]{@{}c@{}}Brightness peak \\ {[}Jy/beam{]}\end{tabular}} & \multicolumn{1}{c|}{\begin{tabular}[c]{@{}c@{}}$\sigma_\mathrm{rms}$\\ {[}mJy/beam{]}\end{tabular}} \\ 

\hline 
\hline

\multicolumn{1}{c}{43.2}  & \multicolumn{1}{c}{MB018A} & \multicolumn{1}{c}{24 Apr 2021}  & \multicolumn{1}{c}{0.343$\times$0.16, 11.6} & \multicolumn{1}{c}{0.340} & \multicolumn{1}{c}{0.162} & \multicolumn{1}{c}{0.125} \\ 

\multicolumn{1}{c}{43.2}  & \multicolumn{1}{c}{MB018B} & \multicolumn{1}{c}{04 Oct 2021}  & \multicolumn{1}{c}{0.461$\times$0.277, - 25.2} & \multicolumn{1}{c}{0.404} & \multicolumn{1}{c}{0.234} & \multicolumn{1}{c}{0.161} \\

\multicolumn{1}{c}{43.2}  & \multicolumn{1}{c}{MB018C} & \multicolumn{1}{c}{02 Apr 2022}  & \multicolumn{1}{c}{0.413$\times$0.184, - 22.1} & \multicolumn{1}{c}{0.413} & \multicolumn{1}{c}{0.231} & \multicolumn{1}{c}{0.043} \\ 

\hline \hline

\multicolumn{1}{c}{1.4}  & \multicolumn{1}{c}{BP243} & \multicolumn{1}{c}{05 Jan 2020}  & \multicolumn{1}{c}{9.806$\times$6.220, - 2.8} & \multicolumn{1}{c}{0.387} & \multicolumn{1}{c}{0.241} & \multicolumn{1}{c}{0.047} \\ 

\multicolumn{1}{c}{2.3}  & & \multicolumn{1}{c}{}  & \multicolumn{1}{c}{6.462$\times$4.790, - 0.8} & \multicolumn{1}{c}{0.454} & \multicolumn{1}{c}{0.291} & \multicolumn{1}{c}{0.196} \\

\multicolumn{1}{c}{5.0}  & & \multicolumn{1}{c}{}  & \multicolumn{1}{c}{2.855$\times$1.350, - 5.1} & \multicolumn{1}{c}{0.594} & \multicolumn{1}{c}{0.360} & \multicolumn{1}{c}{0.041} \\ 

\multicolumn{1}{c}{8.4}  & & \multicolumn{1}{c}{}  & \multicolumn{1}{c}{1.709$\times$1.113, - 10.7} & \multicolumn{1}{c}{0.657} & \multicolumn{1}{c}{0.401} & \multicolumn{1}{c}{0.063} \\

\multicolumn{1}{c}{15.3}  & & \multicolumn{1}{c}{}  & \multicolumn{1}{c}{0.964$\times$0.672, - 9.3} & \multicolumn{1}{c}{0.626} &  \multicolumn{1}{c}{0.311} & \multicolumn{1}{c}{0.067} \\ 

\multicolumn{1}{c}{22.2}  & & \multicolumn{1}{c}{}  & \multicolumn{1}{c}{0.804$\times$0.521, - 9.3} & \multicolumn{1}{c}{0.524} &  \multicolumn{1}{c}{0.267} & \multicolumn{1}{c}{0.096} \\ 

\multicolumn{1}{c}{43.0}  & & \multicolumn{1}{c}{}  & \multicolumn{1}{c}{0.533$\times$0.461, - 9.3} & \multicolumn{1}{c}{0.325} &  \multicolumn{1}{c}{0.252} & \multicolumn{1}{c}{0.542} \\ 

\hline

\end{tabular}
\begin{flushleft} 
\textbf{Notes.} Column 1: observing frequency in\,GHz; Column 2: project code; Column 3: date of the observation; Column 4: beam size and position angle; Column 5: total flux density in Jy; Column 6: brightness peak value in Jy/beam; Column 7: thermal noise in mJy/beam.
\end{flushleft}
\label{table:original_maps}
\end{table*}

In this article, we present and analyze three new 43 GHz VLBA images of NGC\,315.
The experiments were carried out during the GMVA observing sessions in April 2021/April 2022, with a six-month cadence. 
Results from the 86\,GHz observations will be presented in a separate and future work.

The data were calibrated in AIPS \citep{Greisen1990}, following the standard procedure for VLBA data sets.
The imaging, together with the amplitude and phase self-calibrations, was performed using DIFMAP \citep{Shepherd1997}. 
The first and second epochs (April 2021 and October 2021, respectively) suffered from flux density scaling problems at 7\,mm arising, respectively, during the data correlation and due to issues with the focus and rotation at the VLBA stations
\footnote{\url{https://science.nrao.edu/facilities/vlba/data-processing/7mm-performance-2021}}. 
For the April 2021 observations, the problem was solved by applying an antenna-based scaling factor of $\sqrt{2}$, while corrected gain curves provided by the VLBA staff were used for correcting the amplitude calibration of the October 2021 data.

The observation logs and properties of the clean maps produced are reported in Table \ref{table:original_maps}; the clean images are shown in Fig.~\ref{fig:VLBA_43GHz}.
Remarkably, epoch October 2021 shows hints of a stratified jet, with a double ridgeline\footnote{The ridgeline is an indication of the jet direction, and it is defined as the line connecting the brightest pixel at each consecutive slice transversal to the jet direction.}, namely limb-brightening, appearing on scale $\sim 4 \, \mathrm{mas}$ downstream the radio core.

Along with the new observations, we re-analyzed the multi-frequency VLBA data set published by \cite{Park2021}, comprising observations at seven frequencies in the range (1.7-43)\,GHz (Table \ref{table:original_maps}), observed simultaneously.
While a spectral index analysis for such observations was already published \citep{Park2021}, we aim at expanding it (see Sect.~\ref{sec:spectral_index}) by also performing a synchrotron turnover frequency study (see Sect.~\ref{sec:turnover_frequency}).
The obtained clean images are mostly comparable to those reported by \citet{Park2021}.
Based on our assumption of a flux density uncertainty of 5\%, the peak brightness is comparable for all frequencies except at 15\,GHz. Indeed, at this frequency, we recover a brightness peak $\sim 9\%$ higher using the same restoring beam.
The higher flux density of our 15\,GHz map better matches the value expected based on the results obtained at the other frequencies (see Appendix \ref{app:data}). 
After imaging in DIFMAP, the polarization calibration for the 43\ GHz was handled using the ParselTongue-based pipeline GPCAL \citep{Park2021_GPCAL}.

Finally, to increase the number of data points in the brightness temperature profiles (see Sect.~\ref{sec:brightness_temperature}), we further include 15\,GHz observations from the MOJAVE 
\footnote{see \url{https://www.cv.nrao.edu/MOJAVE/sourcepages/0055+300.shtml}}
monitoring program \citep{Lister2018} and VLBI images published by \citet{Boccardi2021}.
From the latter work, we also include 22\,GHz and 43\,GHz observations from 2008 and 2018 which were used by \citet{Ricci2022} to perform a first spectral analysis and are here shown as well to be compared with our new results (Sect.~\ref{sec:spectral_index}).

\subsection{Core shift} \label{sec:core_shift}

Due to synchrotron self-absorption, the optically thick-to-optically thin transition surface appears to move downstream of the jet as the observing frequency decreases \citep{Blandford1979}. 
Therefore, the analysis of a multi-frequency data set requires aligning the maps to a common reference point, which needs to be determined and which lies, ideally, as close as possible to the position of the central black hole. 
In this work, we assume as a reference point the position of the VLBI core at 43\,GHz, as previously done by \cite{Boccardi2021, Ricci2022}.
We discuss the validity of this assumption when exploring the putative position of the central black hole (Sect.~\ref{sec:real_black_hole}).

Several methods have been developed to determine the core shift \citep[see, for example,][]{OSullivan2009, Fromm2013b}.
In this work, we determine the core shift by aligning the optically thin region of the jet at two adjacent observing frequencies, using a 2D cross-correlation. 
The optically thick region is masked during this process and will appear shifted between two frequencies following the alignment.
For each frequency pair, maps are created using the same pixel size and restoring beam, while the self-calibrated data are restricted to the same uv-range\footnote{We performed this analysis also by cutting the uv-range before self-calibration and we obtain comparable results.}. 
We restore the clean maps using the equivalent circular beam of the lower-frequency map, while the pixel size is set to one-tenth of the beam size.
Our results on the core-shift are summarized in Table \ref{tab:core_shift}.
As error, we sum the uncertainties on both axes in quadrature, which are assumed to be equal to the pixel size.
The latter is indeed the minimum displacement that can be computed in the cross-correlation.
Our core-shift estimates confirm those of the apparent core in \citet{Park2021}.

\begin{figure}[t]
    \centering
    \includegraphics[width=\linewidth]{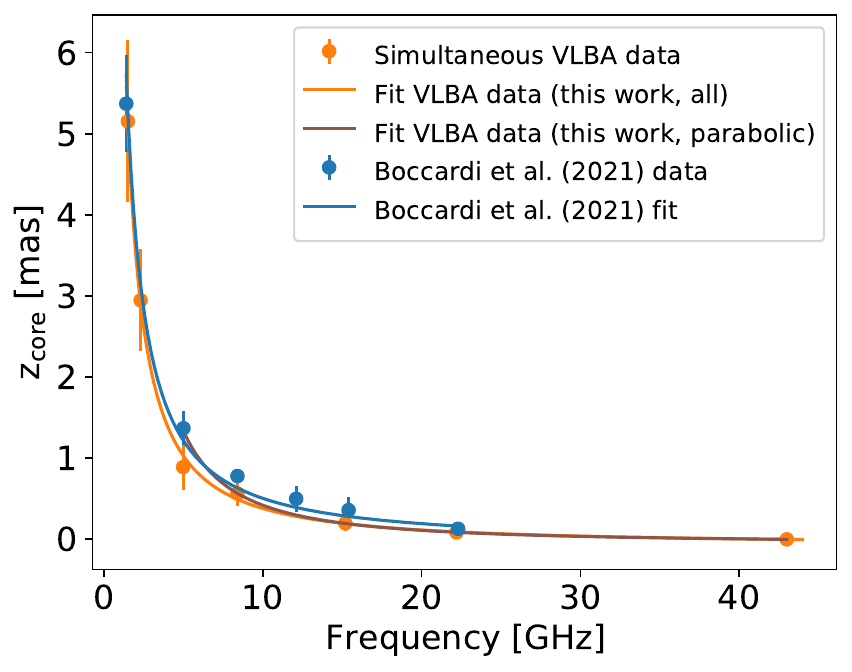}\par
    \caption{Core position as a function of frequency for the two different data sets: i) the multi-frequency VLBA data set (orange points); ii) the multi-frequency and multi-epoch data set presented in \citet{Boccardi2021} (blue points). The blue line represents the best-fit values presented in \citet{Boccardi2021}, the orange line traces the best-fit performed in this paper by employing all the data, and the purple line highlights the best-fit curve obtained using only the data points at 8, 15, 22, and 43\,GHz.}
    \label{fig:core_shift}
\end{figure}

In Fig.~\ref{fig:core_shift}, we show the core shift for two different data sets. 
The orange data points are reported in Table \ref{tab:core_shift}, while the blue ones are taken from \citet{Boccardi2021}.
To explore whether the core shift shows any time variability, we compare our results with theirs since the same method to estimate the shifts has been applied.
We fit our core-shift data with a power-law function in the form $\mathrm{z}_\mathrm{core} = c_s + a \cdot \nu^{1/k_r}$.
The best-fit values are $c_s = - 0.07 \pm 0.03$, $a = 8.7 \pm 0.6$ and $k_r = -0.78 \pm 0.07$ and are shown as the orange continuous line in Fig.\,\ref{fig:core_shift}.

However, as highlighted in \citet{Ricci2022}, the core shift behaves differently between the conical and parabolic regimes in a jet.
Therefore, we also perform the power-law fit using only the points sampling the parabolic region proposed in \citet{Boccardi2021}, namely, 8, 15, 22, and 43\,GHz leading to $c_s = -0.048 \pm 0.005$, $a = 17.0 \pm 1.5$ and $k_r = -0.64 \pm 0.04$ (purple continuous line in Fig.\,\ref{fig:core_shift}).
This latter function is important when calculating the putative position of the central black hole (see Sect.\, \ref{sec:real_black_hole}).
Finally, the blue continuous line represents the best-fit values according to \citet{Boccardi2021} with $k_r = -0.84 \pm 0.06$, consistent within 1$\sigma$ with the value previously reported using all the frequencies.

Throughout the paper, we will use the two different core-shift measurements depending on the observations we are considering: those reported in Table \ref{tab:core_shift} for the VLBA simultaneous data set and those published in \citet{Boccardi2021} for their data set.

\begin{table}[]
\caption{Core-shift measurements relative to the 43\,GHz core for the simultaneous VLBA data set.}
\centering
\begin{tabular}{llll}
\hline
\multicolumn{1}{|c|}{\begin{tabular}[c]{@{}c@{}}$\nu$\\ {[}GHz{]}\end{tabular}} &  \multicolumn{1}{|c|}{\begin{tabular}[c]{@{}c@{}}$\Delta_\mathrm{x}$\\ {[}mas{]}\end{tabular}} & \multicolumn{1}{|c|}{\begin{tabular}[c]{@{}c@{}}$\Delta_\mathrm{y}$\\ {[}mas{]}\end{tabular}} & \multicolumn{1}{|c|}{\begin{tabular}[c]{@{}c@{}}$\Delta_\mathrm{z}$\\ {[}mas{]}\end{tabular}} \\ \hline \hline
\multicolumn{1}{c}{1.4} & $3.90 \pm 1.00$ & $-3.33 \pm 1.00$ & \multicolumn{1}{c}{$5.15 \pm 1.00$} \\ 
\multicolumn{1}{c}{2.3} & $2.34 \pm 0.63$ & $-1.77 \pm 0.63$ & \multicolumn{1}{c}{$2.95 \pm 0.63$}
\\ 
\multicolumn{1}{c}{5.0} & $0.63 \pm 0.28$ & $-0.63 \pm 0.28$ & \multicolumn{1}{c}{$0.89 \pm 0.28$}
\\ 
\multicolumn{1}{c}{8.4} & $0.40 \pm 0.16$ & $-0.40 \pm 0.16$ & \multicolumn{1}{c}{$0.57 \pm 0.16$}
\\
\multicolumn{1}{c}{15.3} & $0.14 \pm 0.10$ & $-0.14 \pm 0.10$ & \multicolumn{1}{c}{$0.20 \pm 0.10$} \\
\multicolumn{1}{c}{22.2} & $0.06 \pm 0.06$ & $-0.06 \pm 0.06$ & \multicolumn{1}{c}{$0.08 \pm 0.06$} \\
\multicolumn{1}{c}{43.0} & \multicolumn{1}{c}{$0.0 $} & \multicolumn{1}{c}{$0.0 $} & \multicolumn{1}{c}{$0.0 $} \\
\hline

\end{tabular}
\begin{flushleft}
\textbf{Notes.} Column 1: observing frequency in GHz; Column 2: displacement along the right ascension in mas; Column 3: displacement along the declination in mas; Column 4: radial displacement in mas.
\end{flushleft}
\label{tab:core_shift}
\end{table}

\subsection{Gaussian modelfit components} \label{sec:modelfit_comp}

To analyze the evolution of the brightness temperature along the jet (see Sect.~\ref{sec:brightness_temperature}) we model the visibilities with Gaussian circular components. 
This procedure is performed using the subroutine \textit{modelfit} in DIFMAP.
The obtained parameters (frequency, flux density, position, and size) of the components are reported in Appendix \ref{app:gaussian_comp}.
We perform such analysis for the new 43\,GHz, as well as for the VLBA simultaneous data set and the 15\,GHz MOJAVE data.
For the maps published in \citet{Boccardi2021}, we use the components they report in their Appendix A.
All the maps are first shifted such that the core component sits at the coordinate origin, and then shifted again according to the core-shift estimates (Sect.~\ref{sec:core_shift}).

To compute the errors on the modelfit parameters, we follow a conservative approach.
The component flux errors are set to 10\%, the uncertainty of the size is assumed to be 25\% of the measured FWHM, and the position errors are computed as the quadrature of 20\% of the radial distance of the component to the core, with the core shift uncertainty added \citep{Boccardi2021}. 
Additionally, we filter out the components whose FWHM size is smaller than 20\% of the minor beam axis.

\section{Results} \label{sec:results}

\subsection{Black hole position} \label{sec:real_black_hole}

\begin{figure*}[t]
    \centering
    \begin{multicols}{2}
    \includegraphics[width=\linewidth]{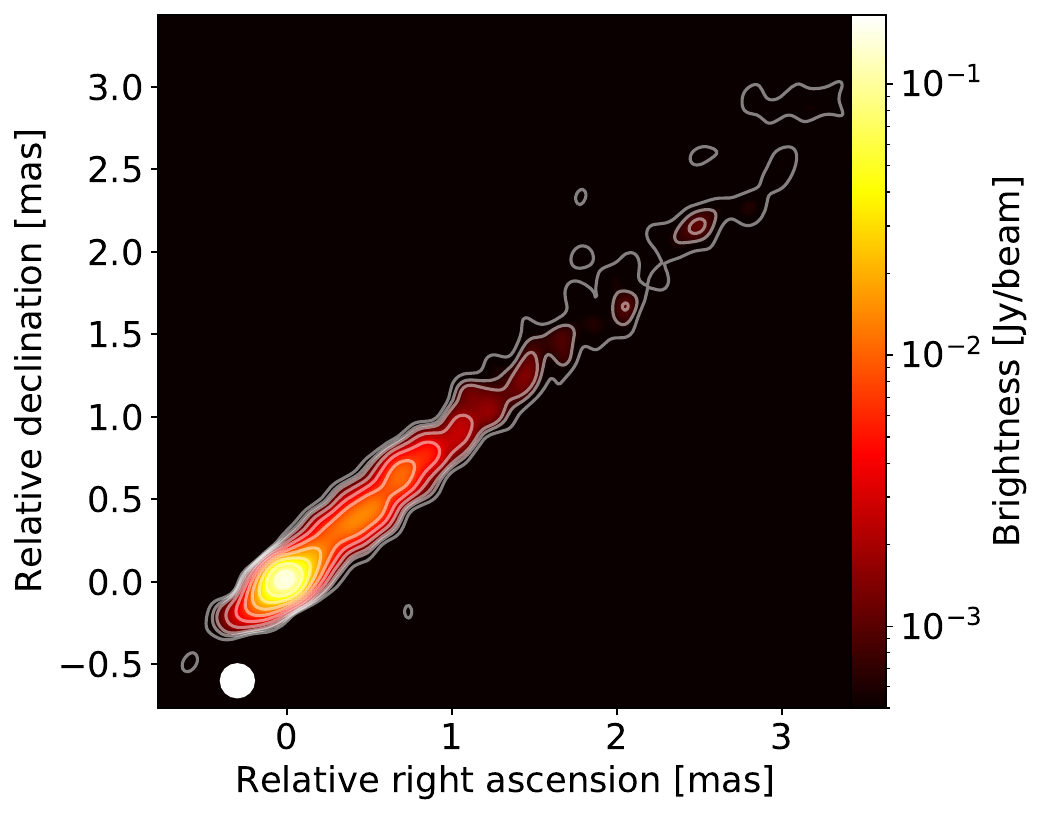}\par
    \includegraphics[width=0.95\linewidth]{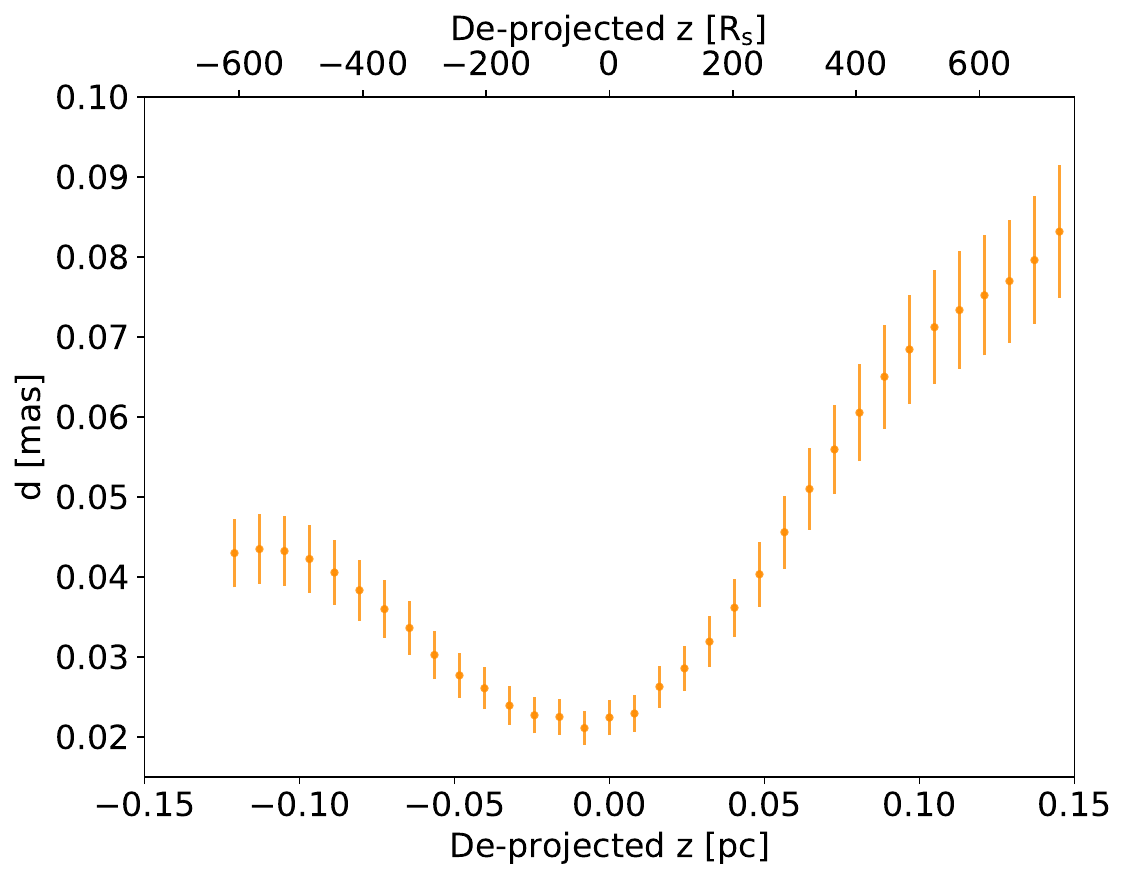}\par
    \end{multicols}
    \caption{Left panel: stacked image at 43\,GHz created using four different epochs, each restored with a common circular beam. Right panel: jet width ($\mathrm{d}$) as a function of the de-projected distance from the core ($\mathrm{z}$), based on the analysis of the super-resolved stacked image. The minimum jet width, observed at $0.008 \pm 0.008 \, \mathrm{pc} = 41 \pm \, 41 \, \mathrm{R_S}$, may pinpoint the position of the black hole.}
    \label{fig:black_hole_pos}
\end{figure*}


In order to attempt to locate the position of the central black hole, we used a 43\,GHz stacked image obtained by combining the three epochs described in Sect.~\ref{sec:original_data} and the 2018 epoch observation published in \citet{Boccardi2021}. 
The maps are uniformly weighted and convolved with 65\% of the average equivalently circular beam, mildly super-resolving the images. 
The resulting map is displayed in Fig.~\ref{fig:black_hole_pos}, left panel. 
From there, we obtained the jet width profile shown in the right panel of Fig.~\ref{fig:black_hole_pos}, employing the pixel-by-pixel code described in \citet{Ricci2022}. 

Given the detection of both the jet and the counter-jet, we identify the position of the minimum of the width with the jet injection point, namely the most likely location of the central black hole in 43\,GHz stacked image.
We find such minimum $0.008 \pm 0.008 \, \mathrm{pc} = 41 \pm \, 41 \, \mathrm{R_s}$ upstream of the brightest pixel at 43\,GHz, in the direction of the counter-jet, as expected due to the core shift. 

A second constraint on the black hole position with respect to the 43\,GHz core can be obtained by extrapolating the core-shift function to infinite frequencies.
Using the best-fit values reported in Sect.\,\ref{sec:core_shift} for the re-analyzed dataset, we retrieve the black hole position to lay $\mathrm{z_{core}}=188\pm43 \, \mathrm{R_s}$ upstream the 43\,GHz core, when using all the data points (orange line in Fig.\,\ref{fig:core_shift}), and $\mathrm{z_{core}} = 132 \pm 15 \, \mathrm{R_s}$ when employing only the parabolic ones (purple line in Fig.\,\ref{fig:core_shift}).
The extrapolations obtained with this method hint the black hole position to be slightly more upstream with respect to the one inferred from Fig.\,\ref{fig:black_hole_pos}, right panel.

Nonetheless, all the methods suggest the position of the black hole to lay around $\sim 100 \, \mathrm{R_s}$ upstream the 43\,GHz core, consistent with previous estimations for radio galaxies of the position of the mm-core \citep[][]{Hada2011, Baczko2022}.
Additionally, the width inferred at injection is $35 \pm \, 4 \, \mathrm{R_s}$, consistent with the one observed at the jet base in M\,87 \citep[about $30-40 \, \mathrm{R_s}$,][]{Lu2023}.


\subsection{Spectral index} \label{sec:spectral_index}

\begin{figure*}[t]
    \centering
\begin{multicols}{3}
    \includegraphics[width=\linewidth]{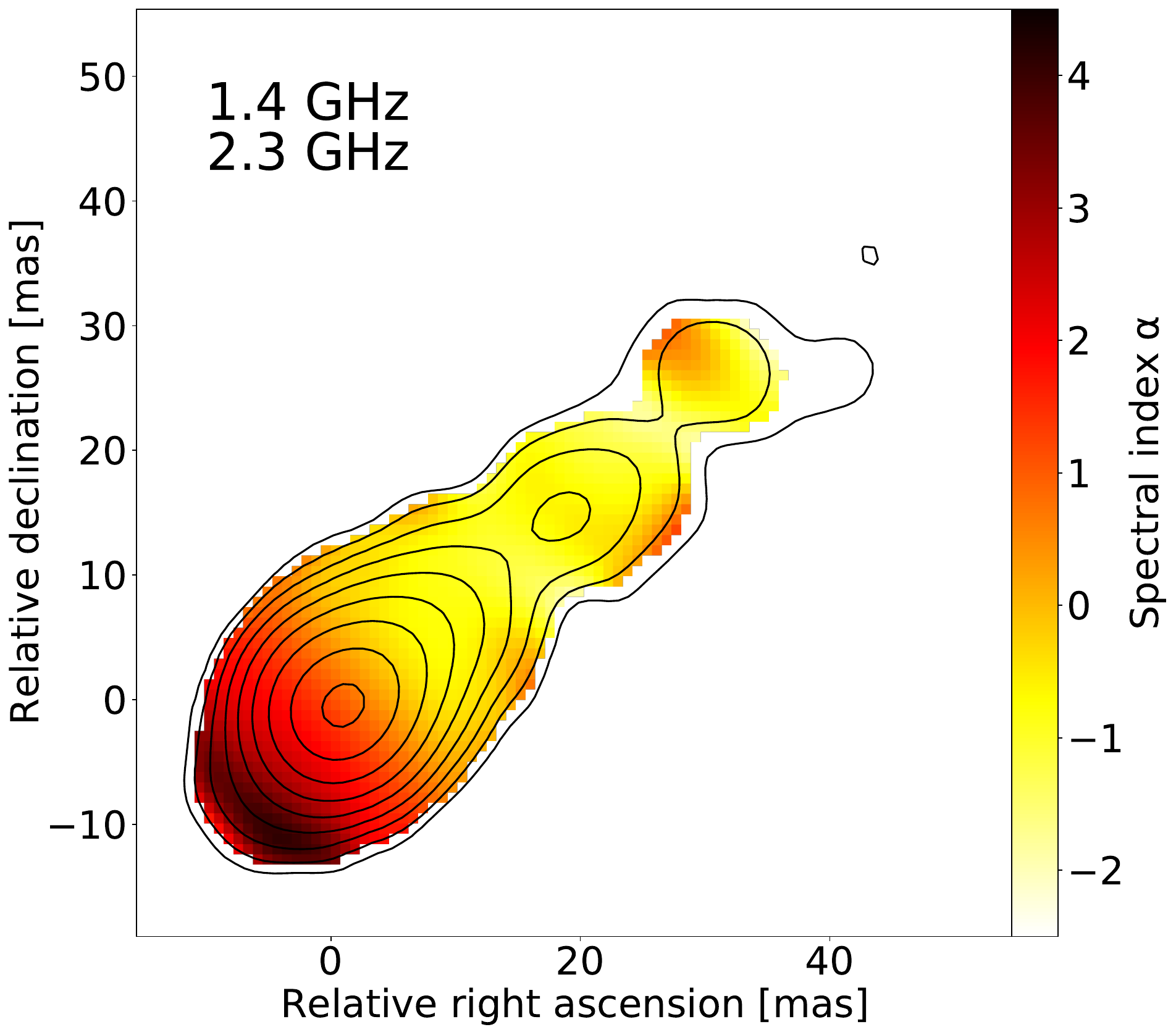}\par
    \includegraphics[width=\linewidth]{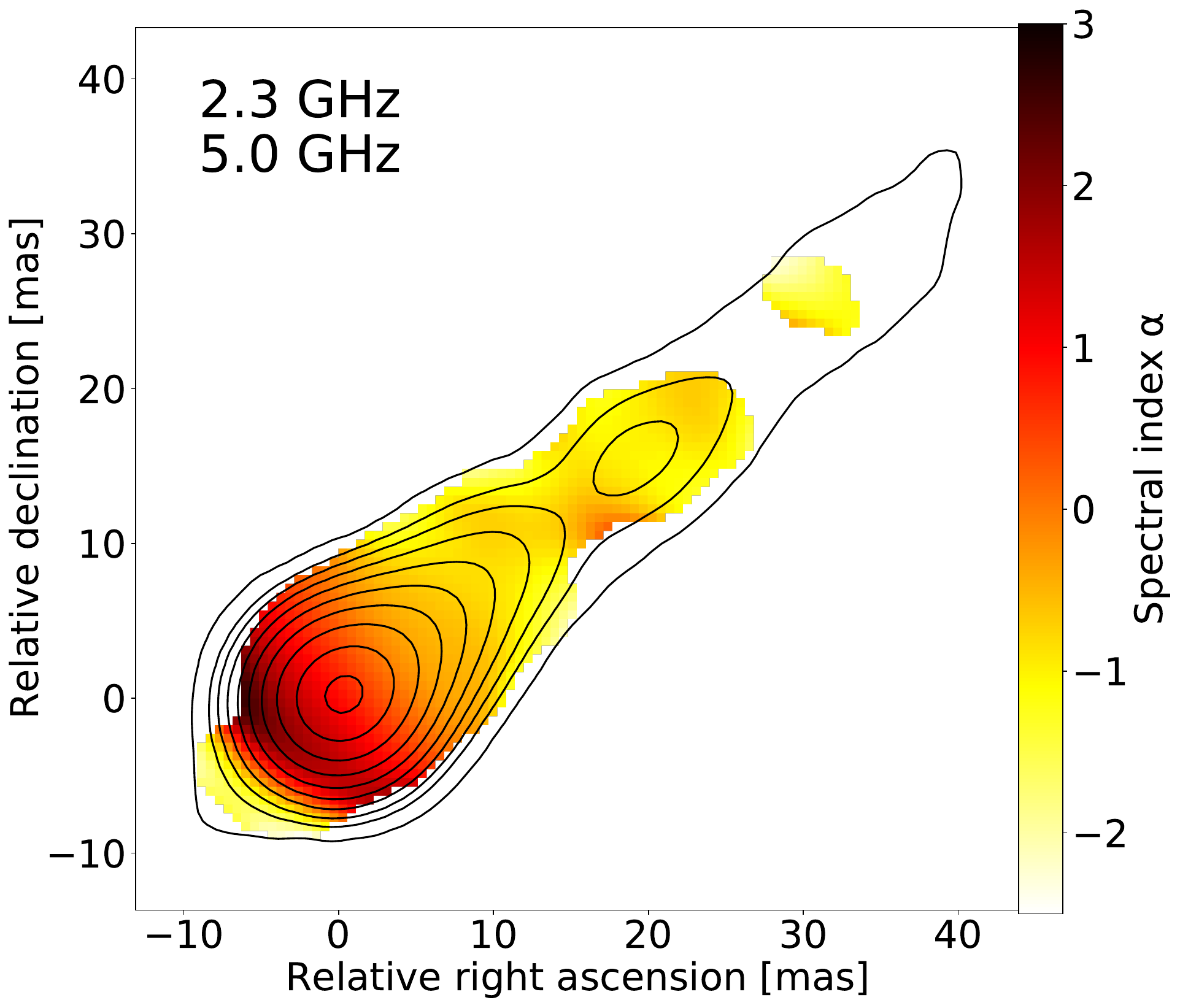}\par
    \includegraphics[width=\linewidth]{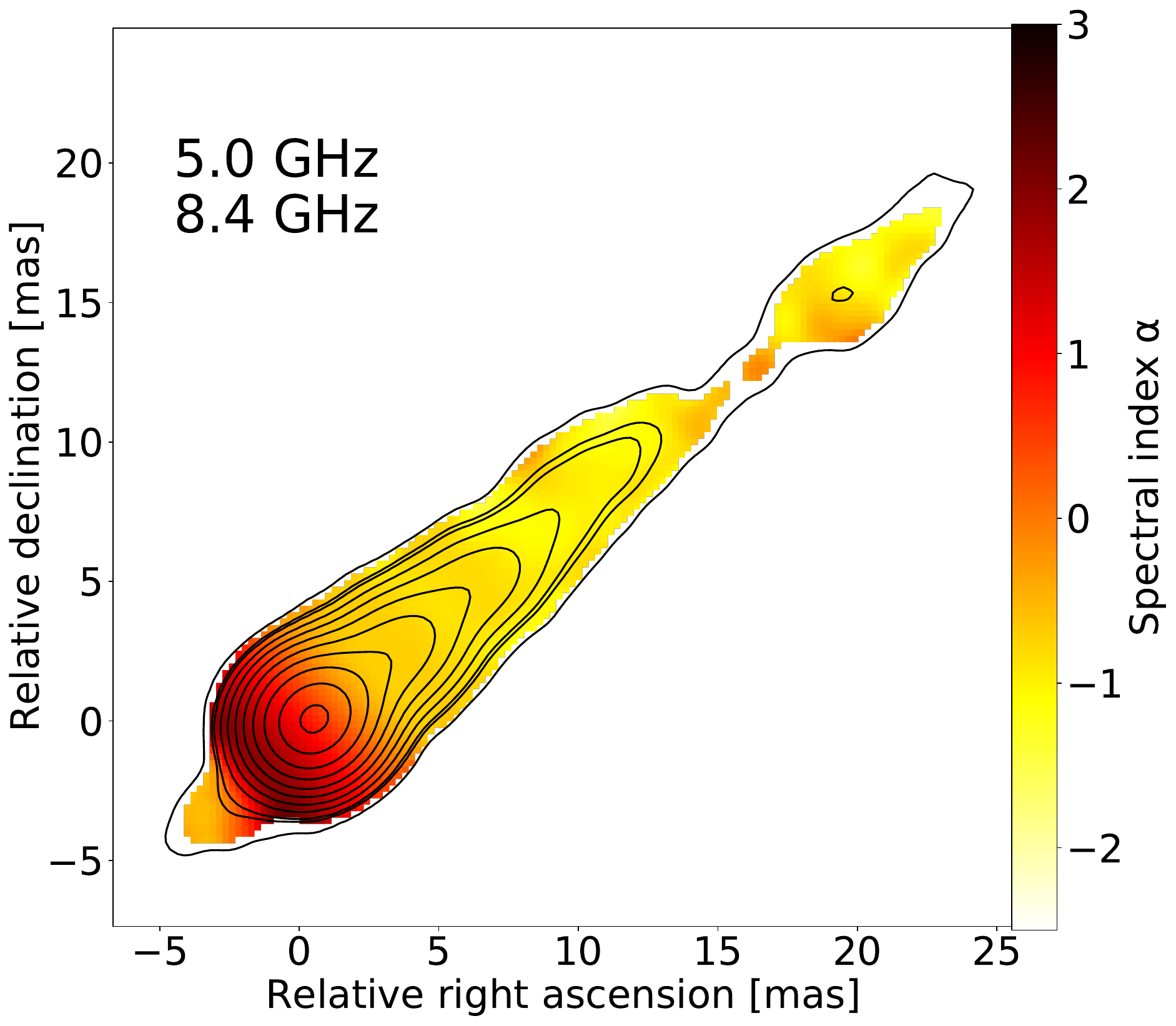}\par
\end{multicols}

\begin{multicols}{3}
    \includegraphics[width=\linewidth]{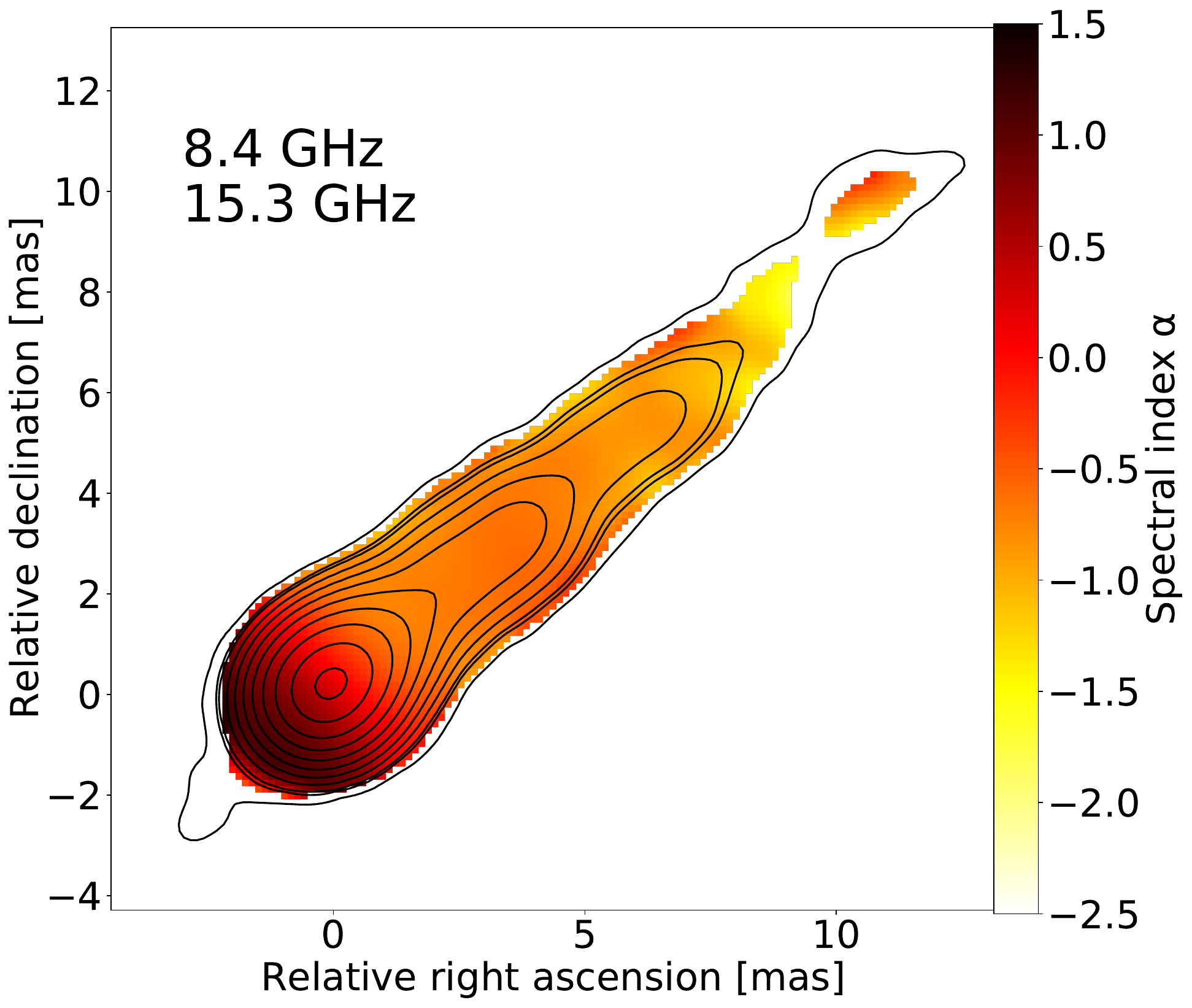}\par
    \includegraphics[width=\linewidth]{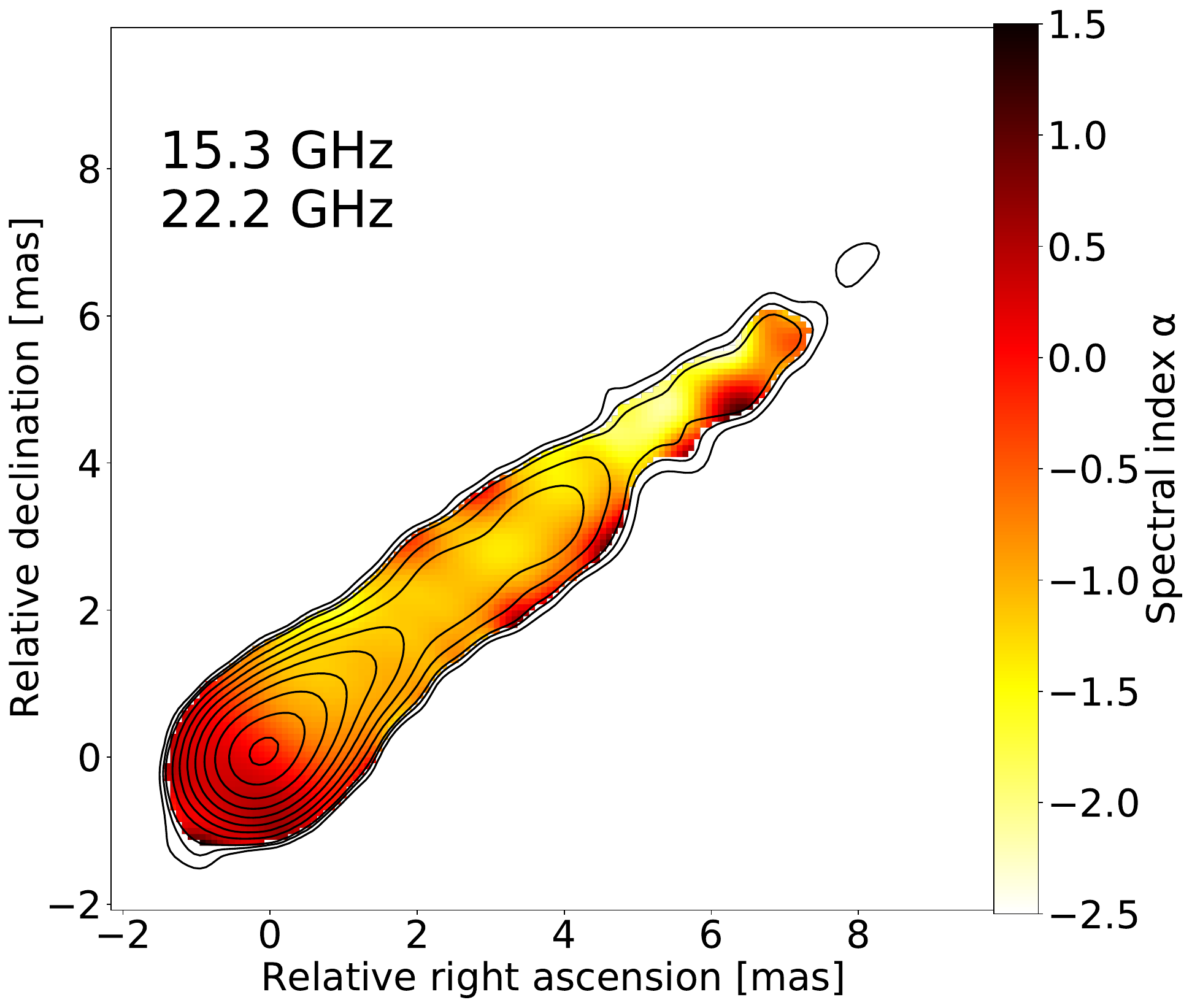} \par
    \includegraphics[width=\linewidth]{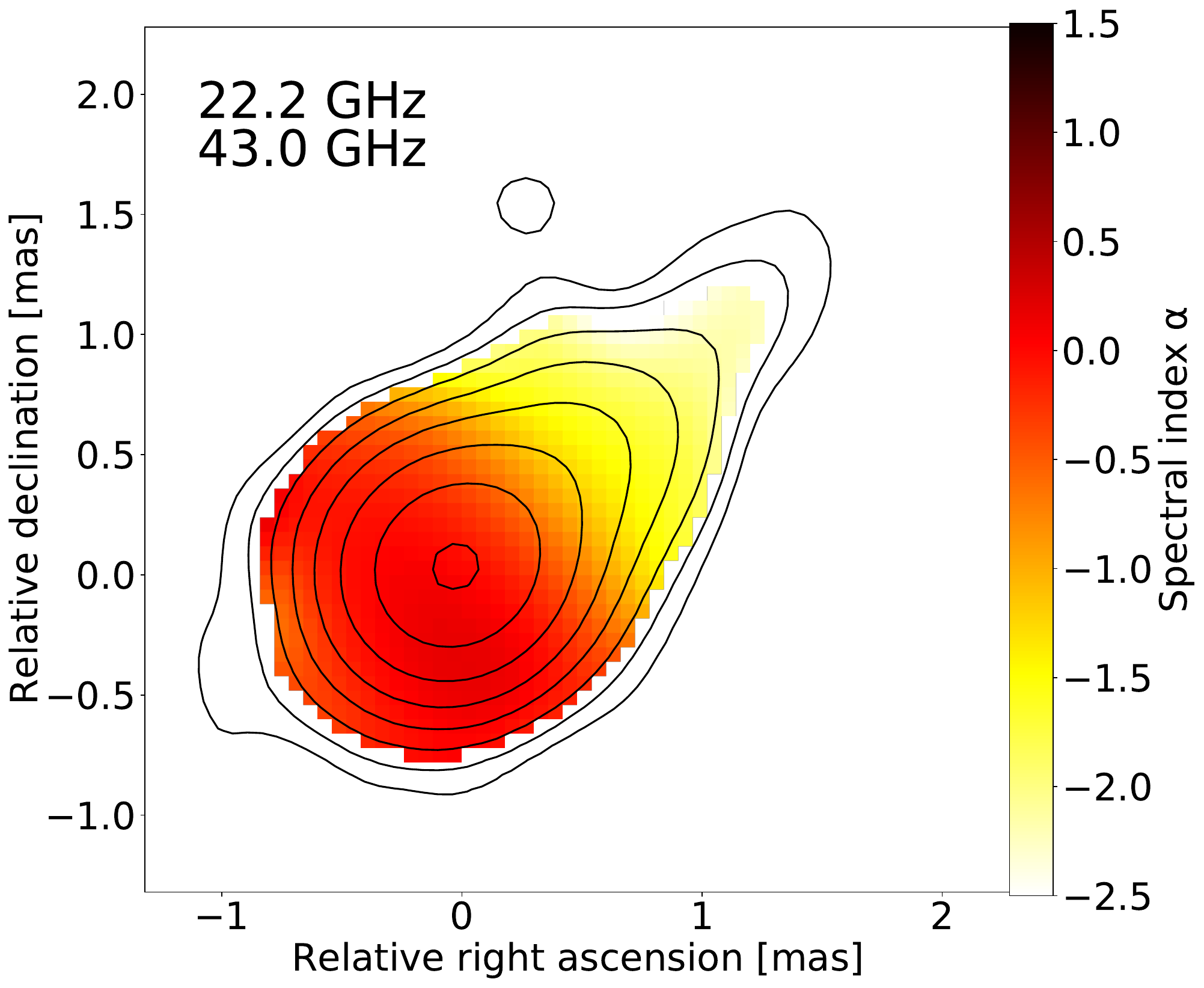}\par
\end{multicols}

    \caption{Spectral index maps between different pairs of frequencies. Top panel starting from the left: (1.4-2.3)\,GHz, (2.3-5.0)\,GHz, and (5.0-8.4)\,GHz. Lower panel starting from the left: (8.4-15.3)\,GHz, (15.3-22.2)\,GHz, and (22.2-43.0)\,GHz.
    The contours trace the total intensity emission in the map at the highest frequency. Note the different color scales.}
    \label{fig:others_alpha_images}
\end{figure*}

Following the procedure described in Sect.~\ref{sec:core_shift} to obtain the core-shift values (Table \ref{tab:core_shift}) and correctly aligning the images, we computed the spectral index maps between pairs of frequencies using the simultaneous observations reported in Table \ref{table:original_maps}.
In the total intensity maps, we filter out pixels with brightness below 10 $\sigma_\mathrm{rms}$.
The spectral maps are presented in Fig.~\ref{fig:others_alpha_images}.

In all maps, we observe a core region characterized by an inverted spectral index, $\alpha > 0$ with $S_{\nu}\propto \nu^{\alpha}$, and a jet with steep to very steep spectral index. 
In two maps, 2--5\,GHz and 5--8\,GHz, the counter-jet is extended enough for the detection of optically thin regions. 
In the former case, the spectral index drops drastically to $\alpha \sim -2$. 
In the latter case, instead, the spectrum is flatter, $\alpha \sim -0.5$, showing similar values with respect to the jet. 
The reason behind this change is still unclear, and a symmetric spectral index gradient in the jet and counter-jet sides is observed only in the 5--8\,GHz case. 

The low-frequency data (1--2\,GHz) indicate the occurrence of a highly-inverted spectrum of $\alpha=4.5$ with $\sigma = 0.30$.
This result is in disagreement with the findings of \citet{Park2021}, in which they recover values up only to $\alpha\sim2.5$ in the nuclear region of the 1--2\,GHz spectral index map.
In the scenario in which the recovered highly-inverted structure is real, it can be seen as an indication of free-free absorption in the nuclear region, highlighting how absorption effects may impact the cm-/mm-VLBI cores.
Indeed, such high spectral index values may be caused by the passage of synchrotron self-absorbed emission with $\alpha=2.5$ through optically thick free-free absorbing material with $\alpha=2$ \citep[see, e.g.,][and references therein]{Kadler2004}.

In the jet region, we observe optically thin synchrotron emission at all frequencies.
In the four lowest frequency maps, the jet steepens reaching the expected values of $-1 \lesssim \alpha \lesssim -0.5$.
On the contrary, the two high-frequency maps reach smaller values of the spectral index, between -1 and -1.5 in the 15-22\,GHz map and down to $\sim - 2.2$ in the 22-43\,GHz map.
Although the 15 and 22\,GHz maps do not give indications of the jet being transversely resolved, we notice that the 15--22\,GHz spectral index map gives indications of a transverse gradient. 
Indeed, along the ridgeline at a distance of about $4 \, \mathrm{mas}$ we observe spectral values $\alpha \sim -1.3$, which rise up to $\alpha \sim -0.5$ towards the edges of the emission.
Hints of a similar structure have also been recovered from the 15--22\,GHz spectral index maps presented in \citet{Park2021, Kino2024}.
Such structure, if real, could be, in principle, connected with an edge-brightened outflow, observed in NGC~315 on similar scales \citep{Park2021}.

\begin{figure*}[t]
    \centering
    \includegraphics[width=0.8\linewidth]{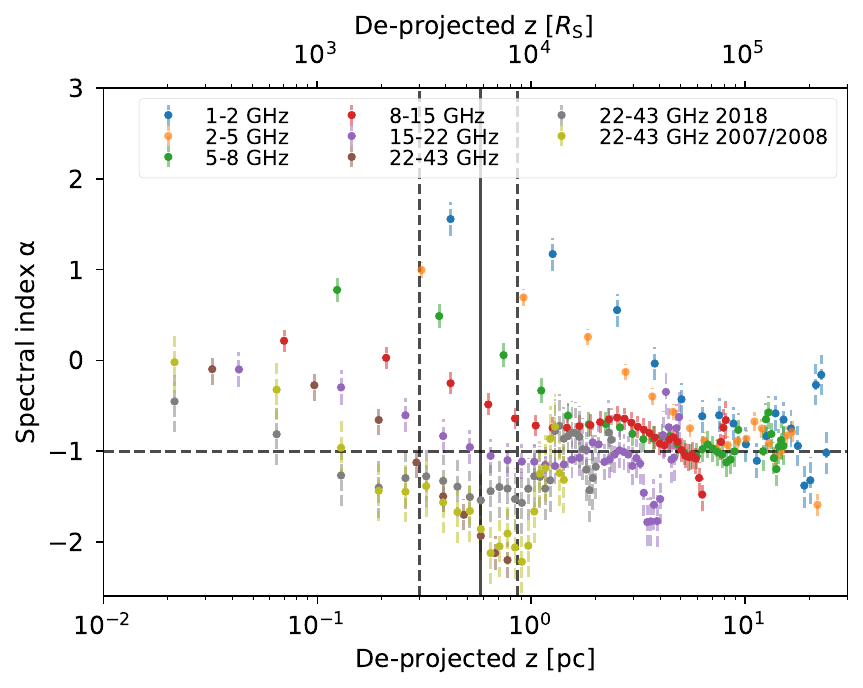}\par
    \caption{Average spectral index as a function of distance from the 43\,GHz core for different pairs of frequencies. In addition to the VLBA data set here presented, we re-present for comparison the 22-43\,GHz spectral index values for the two epochs presented by \citet{Ricci2022}. At high frequencies, remarkably steep spectral index values down to $\alpha \sim -2$ are observed within one parsec from the core, corresponding to ${\sim}10^4 \, R_\mathrm{S}$. Downstream, a convergence towards flatter values $\alpha \sim -0.8$ is observed at all frequencies. The black vertical lines highlight the jet break point, as proposed in \citep{Boccardi2021}, while the horizontal dashed line is set at $\alpha = -1$ as reference.}
    \label{fig:alpha_profiles}
\end{figure*}

In Fig.~\ref{fig:alpha_profiles}, we display the evolution of the average spectral index values along the jet propagation direction.
The positions of the spectral index measurements are shown with respect to the VLBI core at the highest frequency.
The errors are computed by summing in quadrature the thermal noise values together with the flux density uncertainties discussed in Appendix \ref{app:data}.
The very steep spectrum region, observed at high frequencies, has an extent similar to that of the collimation region proposed in \citet{Boccardi2021}.
On parsec scales, all the spectral maps converge towards values of $\sim -0.8$, consistent with the observed radio spectra in the vast majority of AGN jets.
We contextualize these results using two spectral maps between 22 and 43\,GHz obtained in \citet{Ricci2022}.
This comparison offers two different insights: i) it confirms the steep spectrum on sub-parsec scales; ii) it shows that the 22--43\,GHz spectral map obtained in this work is in agreement with the 22--43\,GHz 2007/2008 map, while it shows slightly steeper values than the 22--43\,GHz 2018 counterpart, suggesting possible time-variability of the spectral index.
Possible interpretations of such results are discussed in Sects.~\ref{sec:subparsec_spectral} and \ref{sec:particles}.
Our results are in agreement with the ones inferred for M\,87 by \citet{Ro2023} in two distinct ways.
On the one hand, the authors recover a steep spectral index down to $\sim - 2.5$ on the same scales, namely, $\sim 10^3 \, \mathrm{R_S}$.
On the other hand, the authors show that the spectral index value may change across the different epochs, although not dramatically.
We suggest this variability is likely due to changes in the jet injection conditions and/or in the cooling and re-acceleration mechanisms happening along the jet.

\subsection{Polarization} \label{sec:polarization}

NGC~315 has been found to be unpolarized across a wide range of observing frequencies from 1\,GHz to 43\,GHz \citep{Park2021_GPCAL}.
Nonetheless, we attempt to recover what little polarization is present in the source from our new 43\,GHz data. 
While some weakly polarized features on the mJy-level are occasionally detected, they cannot be recovered or cross-identified in neighboring epochs. 
The lack of jet polarization is a common trait of radio galaxies \citep[see, for example,][]{Pushkarev2017b}.
The underlying cause of the large-scale depolarization, be it the Faraday rotation external and/or internal, beam depolarization, or strong intra-band Faraday rotation, remains unknown \citep{Sokoloff1998,Bower2017, Park2021}.



\subsection{Brightness temperature profile} \label{sec:brightness_temperature}

\begin{figure*}[t]
    \centering
    \begin{multicols}{2}
    \includegraphics[width=\linewidth]{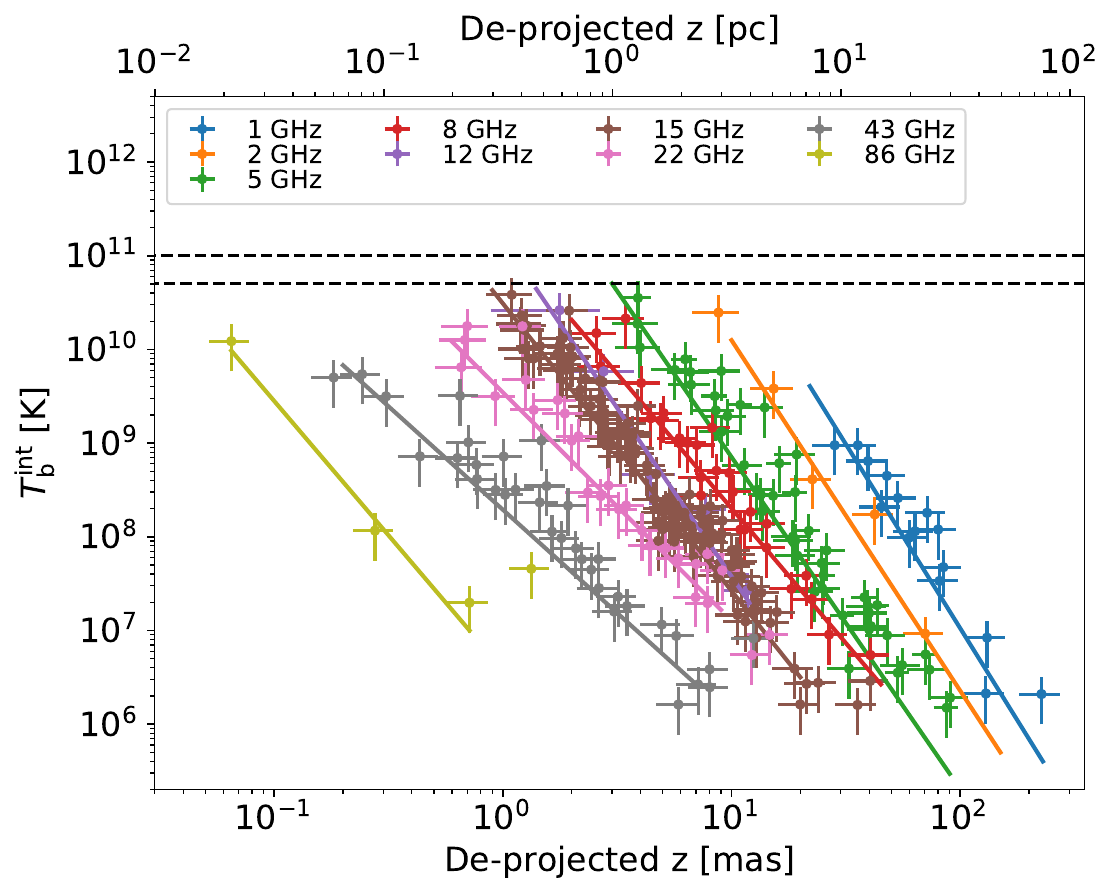}\par
    \includegraphics[width=\linewidth]{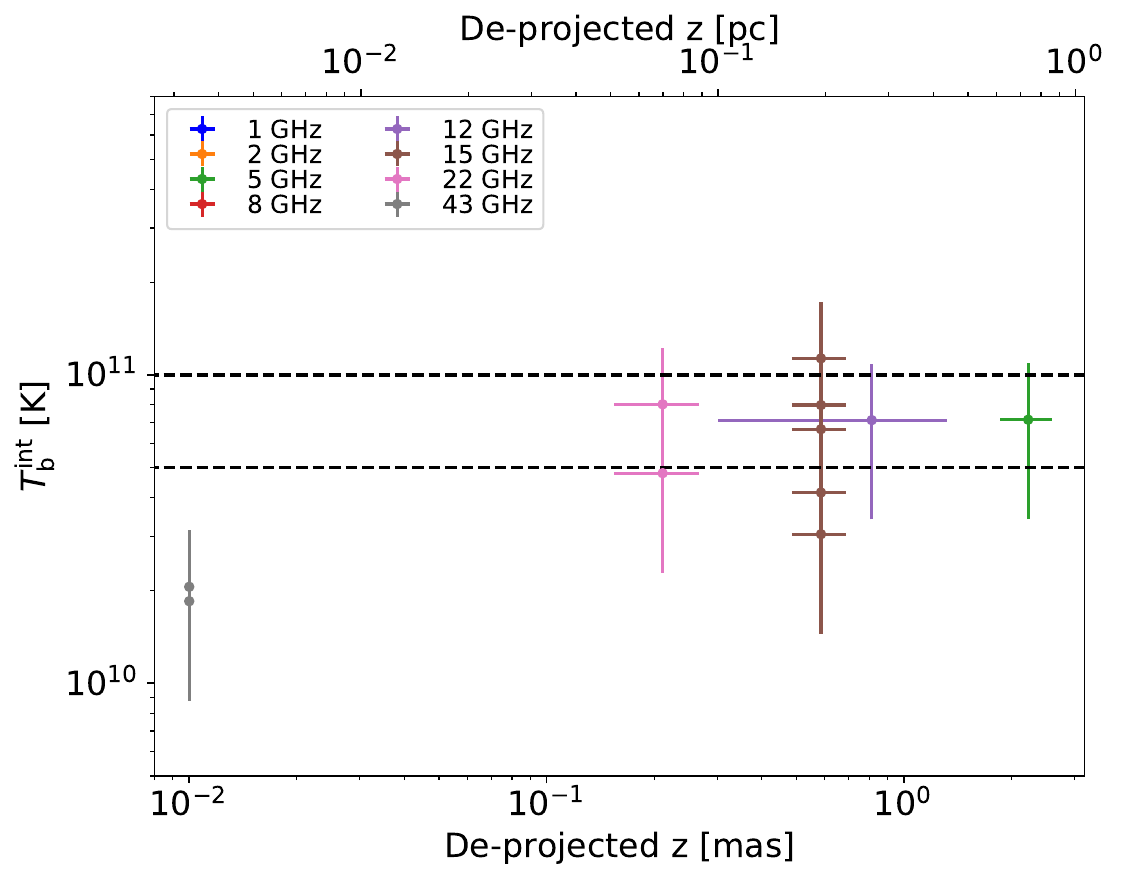}\par
    \end{multicols}
    \caption{Intrinsic brightness temperature profiles of the Gaussian component at different frequencies. The horizontal dashed lines represent the median equipartition brightness temperature derived from sample studies, $T_b = 5 \times 10^{10} \, \mathrm{K}$, and the upper limit $T_b = 10^{11} \, \mathrm{K}$. Left panel: brightness temperature values for the jet components. The continuous lines are the best-fit power laws at different frequencies. Right panel: brightness temperature values for the core components.}
    \label{fig:Tb}
\end{figure*}

In this section, we discuss the intrinsic brightness temperature profiles for the incoming jet of NGC~315.
For a non-thermal source, the apparent brightness temperature is expressed as \citep[see, for example,][]{Kadler2004}:
\begin{equation}
    T_\mathrm{b} = 1.22 \times 10^{12} (1 + z) \Bigg( \frac{S_\nu}{\mathrm{Jy}} \Bigg) \Bigg( \frac{\nu}{\mathrm{GHz}} \Bigg)^{-2} \Bigg( \frac{d}{\mathrm{mas}} \Bigg)^{-2} \, \mathrm{K}  
\end{equation}
in which $S_\nu$ is the flux density, $\nu$ is the frequency and $d$ is the diameter of the emitting region.
The intrinsic brightness temperature is computed as $T_\mathrm{b}^\mathrm{int} = T_\mathrm{b} / \delta$, where $\delta~=~[\Gamma (1 - \beta \mathrm{cos}\theta)]^{-1}$ is the Doppler factor.
The $\beta$ values are taken from the speed profile obtained by \citet{Ricci2022}.
The errors on the intrinsic brightness temperature are computed by propagating the uncertainties on the different parameters defining the Gaussian components described in Sect.~\ref{sec:modelfit_comp}.

In Fig.~\ref{fig:Tb}, we report the intrinsic brightness temperature as a function of distance from the jet apex.
The jet and core components are respectively shown in the left and right panels.
We note that the core components at 43\,GHz should be located on the origin of the x-axis (see Sect.~\ref{sec:core_shift}), but a small shift of 0.01 pc is applied for representing them in a logarithmic scale. 
The data are color-coded by frequency, including results from maps of different epochs.

The dashed horizontal black lines highlight the median and upper limit values of the equipartition brightness temperature.
The latter is achieved when the energy is equally shared between the particles emitting via synchrotron radiation and the magnetic field.
From theoretical and statistical studies, it was shown that the equipartition brightness temperature takes a median value of $T_\mathrm{b} \sim 5 \times 10^{10} \, \mathrm{K}$ \citep[see, for example,][]{Readhead1994, Lahte1999} with the upper limit set to $T_\mathrm{b} \sim 1 \times 10^{11} \, \mathrm{K}$ \citep{Singal2009}.
All jet components fall below the upper limit, while the core components at progressively lower frequencies approach it. 
The inverse Compton catastrophe limit, set around $(5-10) \times 10^{11} \, \mathrm{K}$ \citep{Kellermann1969}, is never exceeded.
Concerning the components associated with the different cores (Fig.\, \ref{fig:Tb}, right panel), the two 43\,GHz points are slightly below the median equipartition brightness temperature.
On the contrary, already on scales of $\sim 0.1 \, \mathrm{pc}$ the quasi totality of the $T_\mathrm{b}$ values of the core components fall in the range of the equipartition state.
Overall, the trend of the core components seems to suggest that on compact scales and in its average state, the jet shows brightness temperatures below the equipartition values, pointing towards a magnetized outflow.
On sub-parsec scales, a flattening around the equipartition values is observed. 
This flattening occurs in proximity to the end of the collimation region, where equipartition conditions are indeed expected.

Along the jet, we fit the brightness temperature profiles with a single power-law $T_\mathrm{b} = T_0 r^\epsilon$. 
For each frequency, the Gaussian components from all the available epochs are fitted jointly.
The best-fit values are reported in Table \ref{tab:Tb}.
The higher frequencies, 22, 43, and 86\,GHz, show a relatively flat slope with $\epsilon \sim -2.2/-2.9$.
On the contrary, at low frequencies, the indices reach values down to $-3.92 \pm 0.30$.
Similarly steep profiles are observed in the eastern jet of NGC~1052 on comparable distances \citep{Kadler2004}.
We highlight that the flatter values are observed at frequencies that sample the parabolic region (or part of it), suggesting different physical conditions with respect to the conical expanding region.
Implications and consequences on the jet nature from such measurements are discussed in Sect.~\ref{sec:jet_evolution}.

\begin{table}[] 

\caption{Power-law indices of the intrinsic brightness temperature profiles at the different frequencies.} 
\centering
\begin{tabular}{c c}
\hline
 \multicolumn{1}{c}{\begin{tabular}[c]{@{}c@{}}$\nu$\\ {[}GHz{]}\end{tabular}} &
 \multicolumn{1}{c}{$\epsilon$} \\

\hline  \hline

1.0 & $-3.92 \pm 0.30$ \\ 
2.0 & $-3.74 \pm 0.38$ \\ 
5.0 & $-3.54 \pm 0.15$ \\
8.4 & $-2.88 \pm 0.13$ \\
12.1 & $-3.59 \pm 0.46$ \\
15.4 & $-3.07 \pm 0.07$ \\
22.3 &  $-2.43 \pm 0.11$ \\
43.2 & $-2.21 \pm 0.10$ \\ 
86.2 & $-2.87 \pm 0.53$ \\ 

\hline

\end{tabular}
\begin{flushleft}
\textbf{Notes.} Column 1: observing frequency in GHz; Column 2: power-law index with the error.
\end{flushleft}
\label{tab:Tb}
\end{table}

\subsection{Turnover frequency} \label{sec:turnover_frequency}

\begin{figure*}[t]
    \centering
    \begin{multicols}{2}
    \includegraphics[width=\linewidth]{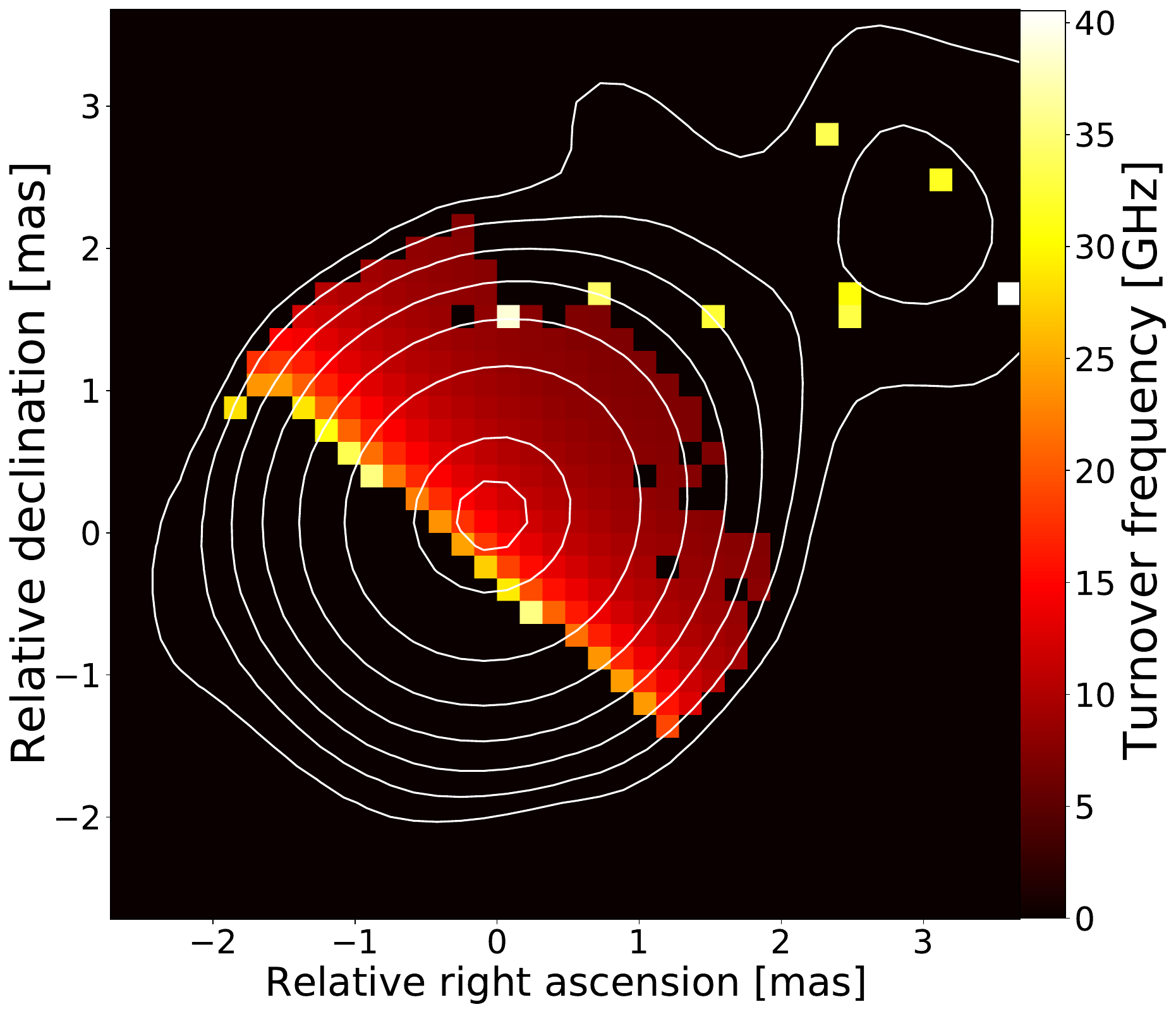}\par
    \includegraphics[width=\linewidth]{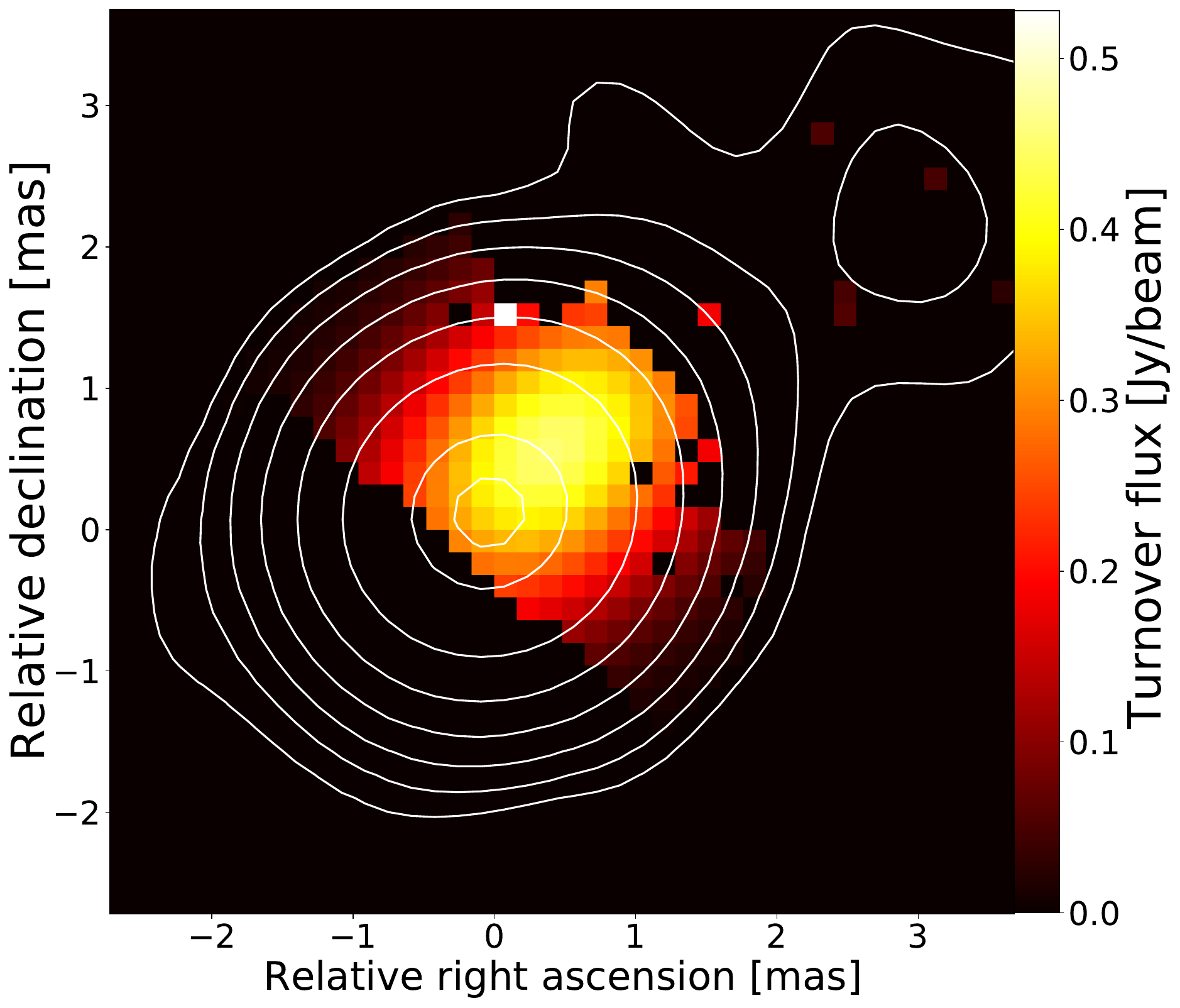}\par
    \end{multicols}
    \caption{Left panel: turnover frequency for each pixel obtained by fitting the synchrotron spectrum across five frequencies between 5 and 43\,GHz. The constant turnover values in the direction perpendicular to the jet propagation point out that the transversal structure is not resolved. The turnover frequency is clearly seen to decrease as going further from the core. The black pixels are spurious data points arising from a not successful fitting of the data. Right panel: brightness at the respective turnover frequency. In both maps, the contours are from the 43\,GHz map.}
    \label{fig:turnover}
\end{figure*}

The turnover frequency of the synchrotron spectrum, namely the transition from an optical thick to an optical thin regime, is related to intrinsic plasma properties, such as the magnetic field strength \citep[see, for example,][]{Lobanov1998}. 
We measured the turnover frequency at each pixel in the core and in the sub-parsec jet region using the simultaneous VLBA observations (see Table \ref{table:original_maps}). 
We use the five highest-frequency VLBA maps convolved with the same beam, pixel size, and uv-range. 
Due to the large difference in resolution between the highest and lowest frequency maps, using all the maps would have led to problems in finding a common beam and pixel size.
For example, including the 1\,GHz and 2\,GHz observations, the common circular beam would have been $\sim 5.08 \, \mathrm{mas}$ and $\sim 3.74 \, \mathrm{mas}$ wide, respectively. 
Convolving the 22 and 43\,GHz maps to such large beams would result in a loss of the majority of the information contained in them.
Consequently, we compute the turnover frequency by using the 5-43\,GHz maps, restored with a common circular beam of $1.65 \, \mathrm{mas}$, larger than $\sim 65\%$ of the beam at 5\,GHz (2.35 mas) and small enough not to lose information in the 22 and 43\,GHz maps.
The turnover frequency for each pixel is obtained by fitting the data points using Eq.~\ref{eq:syn_turn}.
All the parameters vary freely, except for a fixed $\alpha_t=2.5$, following the expectation for a plasma with a homogeneous synchrotron spectrum \citep[][]{Lobanov1998}.
The turnover frequency and flux maps are reported in Fig.~\ref{fig:turnover} left and right panels, respectively.
The pixels in which either the fit diverged or the turnover is outside of the considered frequency range  are set to zero. For example, this is the case for some regions upstream of the jet base, where the spectrum remains inverted up to the highest observing frequency in the data set. 

We found a gradient in the turnover frequency only in the direction of the jet propagation; due to insufficient transverse resolution, we are unable to investigate the presence of a transverse gradient in the turnover frequency profile.

\begin{figure}[t]
    \centering
    \includegraphics[width=\linewidth]{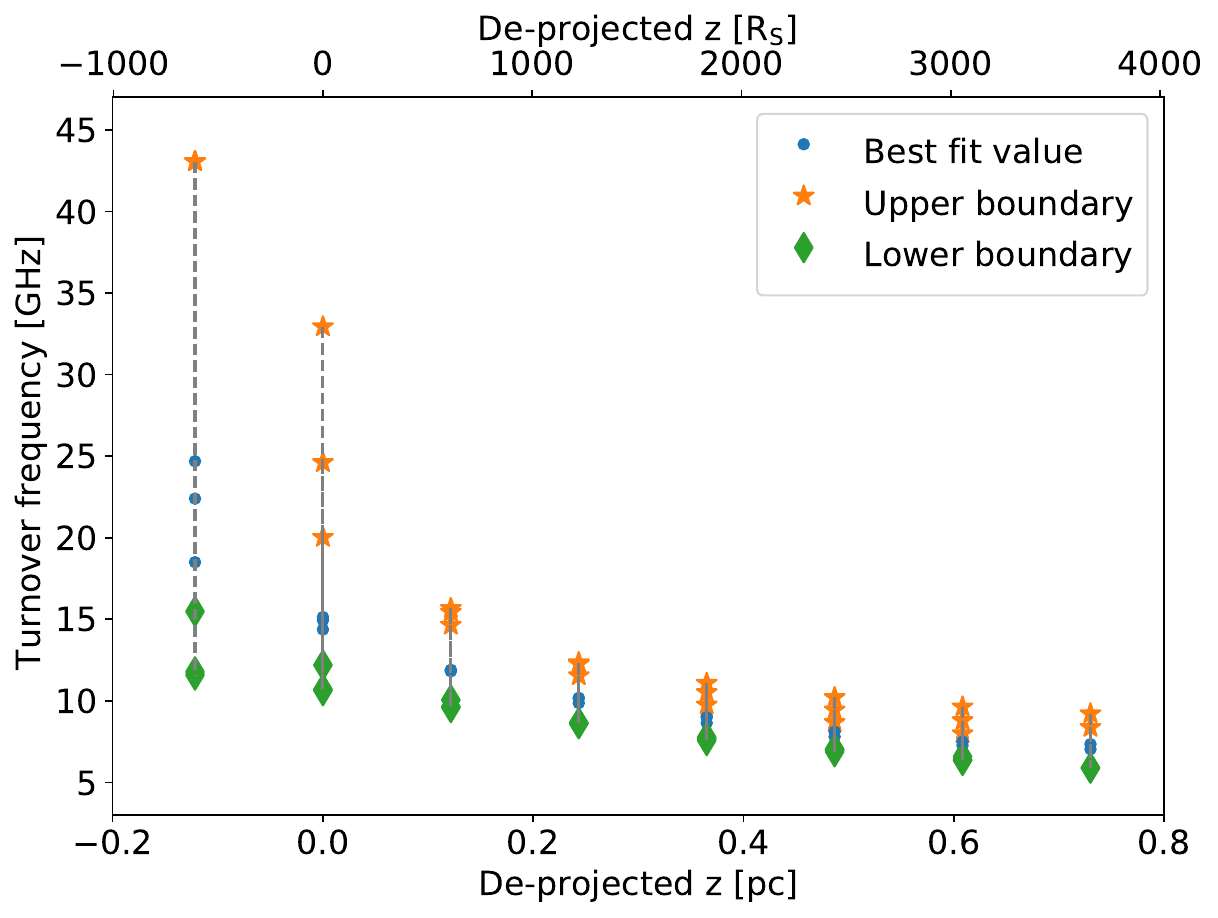}\par 
    \caption{Turnover frequency along the ridgeline as a function of distance from the 43\,GHz core. The multiple data points at the same distance are from the different possible core-shift configurations. The orange points represent the upper boundary, while the green ones the lower boundary. The turnover frequency decreases from $10 \, \mathrm{GHz} \lesssim \nu_\mathrm{br} \lesssim 35 \, \mathrm{GHz}$ in the 43\,GHz core, down to $\sim 6 \, \mathrm{GHz}$ at $\sim 0.7 \, \mathrm{pc}$.}
    \label{fig:turnover_evolution}
\end{figure}

In Fig.~\ref{fig:turnover_evolution}, we show the turnover frequency along the ridgeline as a function of the distance from the 43\,GHz core.
Because of the importance of a proper core shift across all the five frequencies employed, we create a number of turnover frequency maps according to all the possible frequency-displacement configurations.
In detail, with the pixel size of the turnover frequency map being 0.160 mas, different displacements are possible within the estimated errors (see Table \ref{tab:core_shift}).
As highlighted in Fig.~\ref{fig:turnover_evolution}, all the possible turnover frequency maps lead to a turnover frequency of
$10 \, \mathrm{GHz} \lesssim \nu_\mathrm{br} \lesssim 35 \, \mathrm{GHz}$ at the position of the 43\,GHz core and down to $\sim 6 \, \mathrm{GHz}$ at around 0.7 pc.

\section{Discussion} \label{sec:discussion}

\subsection{Steepness of the spectrum} \label{sec:subparsec_spectral}

The variation of the spectral index observed in  Fig.~\ref{fig:alpha_profiles} indicates a change in the physical conditions along the jet.

Models have been developed to explain the different possible spectral index values \citep[see, for example,][]{Hovatta2014, Ro2023}.
The synchrotron spectrum is regulated by the injection of non-thermal particles as a function of the distance from the core, which can be expressed as $Q \propto \mathrm{z}^{q}$. 
Here, $q$ defines the energy spectrum of the injection function for the non-thermal electrons.
When $q = 0$, particles are injected continuously so that the total number of non-thermal particles is conserved along the jet, whereas for $q \rightarrow - \infty$, particles are only injected once at the launching site.
Intermediate values of $q$ indicate the ejection of new electrons along the jet but with their volume decreasing with the distance from the core.
In the $q = 0$ scenario, the spectrum only slightly steepens with the distance.
In contrast, when $q \rightarrow - \infty$, the spectral index drops abruptly and almost immediately to arbitrarily negative low values.
Finally, intermediate values ($-10 \lesssim q \lesssim -5$) describe the spectral indices between 22--43\,GHz, namely in the range $-2.5 \lesssim \alpha \lesssim -1$.
A representation of the spatial evolution of the spectral index downstream of the jet for different values of $q$ is presented by \citet{Ro2023}, in Fig.\ 3.

This model has been applied for NGC\,315 by \citet{Kino2024}. 
By employing the spectral index values obtained from a 15-22\,GHz spectral index map, they constrained values of $-5 \lesssim q \lesssim -2$.
However, our 22--43\,GHz steep spectral indexes (Fig.\, \ref{fig:alpha_profiles}) cannot be described by such small values of $q$, and values $q \sim -10$ better describe the steeper spectrum ($\alpha \sim -2$) \citep[see][]{Ro2023}\footnote{It is important to highlight that the model shown in \citep{Ro2023} was developed for M\,87. However, the two jets, while presenting some differences, are largely similar.}.
Therefore, we propose that on distances of $\sim 10^3 \, \mathrm{R_S}$ a larger value of $q$ is needed, while at about $10^4 \, \mathrm{R_S}$ from the core, namely, in the conical region, the flatter spectrum ($\alpha \sim -0.8$) is better described by values of $q$ closer to zero, implying a higher rate of injection of non-thermal particles, in agreement with what is proposed in \citet{Kino2024}.


As previously highlighted, the steep spectrum region roughly matches the extent of the acceleration and collimation region (see Fig.\ \ref{fig:others_alpha_images}), suggesting that the two phases of the jet geometry are accompanied by differences in the underlying physical processes taking place in the plasma flow.
Such an analysis hints that, on sub-parsec scales, the injection of non-thermal energetic particles decreases with distance from the core, suggesting particle acceleration mechanisms with low efficiency.
In addition, cooling processes could also be important in this region.
The cooling can, for instance, be explained by synchrotron losses due to the strong magnetic fields dominating on such scales \citep{Kardashev1962, Pacho1970, Blandford1979}, a phenomenon already proposed to be happening on the same scales in NGC\,315 \citep{Ricci2022} and M\,87 \citep{Nikonov2023}.
On the contrary, on parsec scales, we observe re-heating and re-energizing of new particles, which could be a consequence of more efficient particle acceleration mechanisms supported by the presence single or multiple standing shocks, such as reconfinement shocks.
Additionally, cooling of particles due to synchrotron losses is expected to have a weaker, if not null, impact on such scales, due to the decreasing strength of the magnetic field along the jet.
Hints of the presence of recollimation shocks along the jet can be extrapolated by the presence of multiple quasi-stationary features detected by the MOJAVE team at 15\,GHz \citep{Lister2019}\footnote{\url{https://www.cv.nrao.edu/MOJAVE/sepvstime/0055+300_sepvstime.png}}.
In the next section, we discuss whether different particle acceleration mechanisms can account for the two particle injection regimes.

\subsection{Particle acceleration mechanisms} \label{sec:particles}

The evolution of the spectral index from the sub-parsec to the parsec scales raises questions concerning the occurrence of possibly distinct particle acceleration mechanisms.
For the sake of simplicity, we put the majority of the focus on two of the most common acceleration processes in relativistic objects: diffusive shock acceleration (hereafter DSA, \citealt{Blandford1987}) and magnetic reconnection \citep{Loureiro2016}.
As a first consideration, we point out that these two acceleration processes take place in different regimes.
For instance, in a strongly magnetized plasma, magnetic reconnection is able to accelerate particles more efficiently, while the DSA is suppressed.
Conversely, the reconnection process is less likely to take place in weakly magnetized plasmas \citep{Ripperda2019}.

On sub-parsec scales, the spectral index steepens to $\alpha \sim -1.5 / -2$, implying a particle energy spectrum of $p_\mathrm{inj} \sim 4$ and $p_\mathrm{inj} \sim 5$, respectively (with $p_\mathrm{inj} = 1 - 2 \alpha$).
Although relativistic magnetic reconnection (occurring below the fluid bulk motion scale) \citep{Sironi2014} is able, in principle, to give rise to such a spectrum, this would be possible only in a relatively weak plasma magnetization regime.
At the same time, as mentioned before in this section, a weak magnetization does not favor the formation of plasmoid instabilities during the reconnection, suppressing, therefore, the efficiency of the process. 
In contrast, such a steep energy spectral index potentially shows a better agreement with the DSA mechanism \citep{Sironi2011} in a moderately magnetized plasma \citep[a magnetized plasma is expected on sub-parsec scales, see][]{Ricci2022, Ricci2024}, with further steepening of the spectrum due to synchrotron losses (see previous section).
In particular, the index of the particle energy spectrum can steepen up to $p_\mathrm{inj} \sim 4.5$ due to the DSA mechanism in case of no particle cooling \citep{Sironi2011}.
Overall on sub-parsec scales, DSA seems to be the most promising mechanism to accelerate particles beyond the thermal distribution in such a source, although magnetic reconnection cannot be ruled out.

Zooming out to the parsec scale, the spectral index ($\alpha\sim-0.8$) is compatible with both DSA \citep{Sironi2009} and magnetic reconnection \citep{Sironi2016}.
However, since magnetic reconnection may be suppressed already at sub-parsec, the DSA mechanism may still be dominant due to the decreasing magnetic field strength.
Additionally, in the scenario the steep-spine/flat-sheath structure observed in the spectral index map 15--22\,GHz is real (Fig.\,\ref{fig:others_alpha_images}), it may be a signature of an efficient shear layer acceleration on parsec scales \citep[see, for example,][and references therein]{Rieger2018, Rieger2019}.
Although such a mechanism is likely not able to account for the steep spectral index values recovered on sub-parsec scales, it may act on parsec scales alongside the DSA.

In conclusion, we find that the spectral index does not completely rule out a specific acceleration process, although DSA seems to be a more realistic process on the scales investigated here. 
Therefore, we propose that the two regimes observed in the spectral index may primarily arise from DSA operating with different efficiencies, due to the varying magnetic field strengths in the two regions, with the additional contribution of synchrotron losses on sub-parsec scales (as discussed in the previous section) and efficient shear layer acceleration on the parsec ones.
Moreover, on parsec scales, the efficiency of the DSA mechanism might be enhanced by the presence of the proposed recollimation shocks (see Sect.\,\ref{sec:subparsec_spectral}) \citep[see, for example,][]{Fuentes2018}.
At smaller distances from the jet launching site, which are not resolved in our observations, we expect reconnection to be dominant, due to the strong magnetic fields.

We point out that our investigation of the particle acceleration processes, which is based purely on the relation between the energy spectrum of the particles and the spectral index due to synchrotron emission, allows us only to hint at the most prominent acceleration processes in each zone and cannot describe the interaction between the different acceleration mechanisms.
Moreover, the scales at which particles accelerate are too small to be captured by both observations and relativistic magnetohydrodynamics simulations, particularly in the magnetic reconnection scenario.
Although we expect reconnection and shocks on the plasma turbulence scales to play a significant role in the large-scale spectrum, their interplay (which would span over $\sim8-10$ order of magnitude in length scales) is still under debate.
High-resolution multi-scale simulations, combined with post-processing techniques (see, for example, \citealt{Seo2023,Meenakhsi2023}) should be able to unveil the true nature of non-thermal particles in astrophysical jets, although, at the current stage and for the upcoming years, the scale separation between particle acceleration and jet propagation is still a critical issue.

\subsection{Magnetic field geometry} \label{sec:jet_evolution}

The VLBI observations of NGC~315 performed thus far indicate that the jet starts off magnetized on compact scales, reaching a moderate magnetization on sub-parsec scales \citep{Ricci2022}.
In the following, we extract further information on the evolution of the magnetic field from the brightness temperature and synchrotron spectrum along the jet, starting from the conical-expanding region.

The slope of the brightness temperature profile ($\epsilon$) is related to the transverse expansion of the components along the jet and to the dominant radiative loss mechanism. 
Specifically, a newly ejected jet component will undergo stages of Compton ($\epsilon_c$), synchrotron ($\epsilon_s$), and adiabatic losses ($\epsilon_a$) \citep{Blandford1979, Marscher1985, Fromm2015}.
Under the assumption of a shock propagating adiabatically downstream a conical jet with constant Lorentz factor, the slope of the brightness temperature profile is, respectively in the three different stages \citep{Lobanov1999, Schinzel2012}:
\begin{equation}
    \epsilon_c = - \left[ \, (|p_\mathrm{inj}| + 5) + |a| \,(|p_\mathrm{inj}| + 1) \, \right] \, / \, 8
    \label{eq:Tbevolution1}
\end{equation}
\begin{equation}
    \epsilon_s = - \left[ \, 4(|p_\mathrm{inj}| + 2) + 3|a| \, (|p_\mathrm{inj}| + 1) \, \right] \, / \, 6
\end{equation}
\begin{equation}
    \epsilon_a = - \left[ \, 2(2|p_\mathrm{inj}| + 1) + 3|a| \, (|p_\mathrm{inj}| + 1 ) \, \right] \, / \, 6
    \label{eq:Tbevolution2}
\end{equation}
Here, the power-law index $a$ defines the magnetic field strength profile ($B(\mathrm{z}) = \mathrm{z}^a$).
\begin{table}[]
\caption{Expected power-law index for the brightness temperature in the jet using the model by \citet{Lobanov1999}.}
\centering
\begin{tabular}{c|cc}
\hline
& \multicolumn{1}{c|}{a = -1} & a = -2           \\ \hline \hline
$\epsilon_c$ & $- 1.40 \pm 0.30$            & $- 1.85 \pm 0.38$ \\ 
$\epsilon_s$ & $- 4.87 \pm 0.68$            & $- 6.67 \pm 0.89$ \\ 
$\epsilon_a$ & $- 3.87 \pm 0.45$            & $- 5.67 \pm 0.67$ \\ \hline
\end{tabular}
\label{tab:losses}
\begin{flushleft}
\textbf{Notes.} Column 1: expected power-law index considering a toroidal field; Column 2: expected power-law index considering a poloidal field.
\end{flushleft}
\end{table}
By assuming an average value for the spectral index (on parsec scales) of $\alpha = - 0.8 \pm 0.3$, with the error chosen to account for the observed fluctuations around $- 0.8$ (see Fig.~\ref{fig:alpha_profiles}), we predict the power-law slopes for the different types of radiative losses for two magnetic field geometries, namely, toroidal ($a = -1$) and poloidal ($a = -2$). 
We note that such magnetic field exponents are valid only when considering a jet expanding with a conical shape.
Indeed the toroidal magnetic field component has a power-law index equal to the opposite of the jet expansion one, namely $-p$ ($d(\mathrm{z}) = \mathrm{z}^p$), while the index of the poloidal component is $-2p$.
The results are reported in Table \ref{tab:losses}.
By comparing these findings with the values reported in Table \ref{tab:Tb}, assuming that the jet components describe the propagation of shocks downstream the outflow, we derive that the fit to the data on parsec scales (Fig.~\ref{fig:Tb}, left panel) is compatible with a jet dominated by adiabatic losses and threaded by a toroidal field.

To explore the magnetic field in the parabolic region we follow the approach proposed by \citet{Kadler2004}.
Assuming 
a power-law for the electron density as $N \propto \mathrm{z}^n$, the power-law index of the brightness temperature is
\begin{equation}
    \epsilon = p + n + a(1 - \alpha) \, .
    \label{eq:B_field_subparsec}
\end{equation}
This relation, while it allows us to consider a non-conical jet shape, is only valid under the assumption of constant Lorentz factor. 
This condition is not strictly valid for NGC~315, although the measured Lorentz factor does not vary by a large factor in the parabolic region \citep{Park2021, Ricci2022}. 
Therefore, we adopt this approach to provide an order-of-magnitude estimate of the magnetic field strength from the jet base down to $\sim 0.6 \, \mathrm{pc}$.
We assume: i) an average spectral index for the sub-parsec scales of $\alpha = -1.80 \pm 0.30$ (Sect.~\ref{sec:spectral_index}), ii) the jet width index for the parabolic region as $p = 0.45 \pm 0.15$ \citep{Boccardi2021}, iii) the brightness temperature index as $\epsilon = - 2.4 \pm 0.5$ as representative of the 22, 43, and 86\,GHz profiles (Sect.~\ref{sec:brightness_temperature}).
For the particle distribution, we consider two possible index values: $n = -1$, a classical assumption for jet evolution \citep{Kadler2004}, and $n = -2$, which describes a particle population without cooling effects.
The inferred index for the magnetic field on the sub-parsec region is $-0.66 \pm 0.14 < a < -0.30 \pm 0.14$.
While the uncertainties associated with this method prevent us from obtaining a precise determination of the slope, Eq.~\ref{eq:B_field_subparsec} indicates that the magnetic field evolves with an index smaller than one within $\sim 10^3 \, \mathrm{R_S}$ from the jet injection point.
Similar results have been obtained for M~87, since \citet{Ro2023} derived an index of $\sim - 0.72$ on the same scales.
In the jet collimation region of NGC~315, the toroidal component evolves as $a =-p=-0.45$, while the poloidal one as $a=-2p=-0.90$. 
However, a steeper evolution for the toroidal field is expected in reality due to its dissipation in the acceleration process \citep[see, for example,][]{Komissarov2007, Komissarov2012}.
This is likely the case for NGC\,315, in which magnetic acceleration throughout the collimation region has been proposed in previous works \citep{Ricci2022, Ricci2024}.
In this scenario, by merging the theoretical expectations from magnetic acceleration and the constraints provided in this section, the magnetic field index in the collimation and acceleration region is expected to lay $-0.80 \lesssim a \lesssim -0.45$.
Ultimately, from an index of $a \lesssim -0.80$, while it is not possible to completely rule out a contribution of the poloidal component, the toroidal magnetic field component is favored.
\citet{Kino2024} from the relation $B(\mathrm{z}) \propto (\mathrm{d} \cdot \Gamma)^{-1}$ (where $\Gamma$ is the Lorentz factor) derive a magnetic field index of $a = -0.88$ for their proposed collimating region.
This value, while slightly outside the range proposed in this work, strengthens the idea that on scales smaller than $\sim 10^4 \mathrm{R_s}$ the magnetic field evolves with an index smaller than one.

Finally, a further investigation of the magnetic field properties on the sub-parsec scales can be based on our previous analysis of the synchrotron turnover frequency. 
Following \citep{Cawthorne1991}:
\begin{equation}
    B(d) = C_0 \nu_\mathrm{br}^5 d^4 I_\mathrm{0}^{-2} 
\label{eq:B_field_turnover}
\end{equation}
where $C_0$ is the normalization factor, and $I_\mathrm{0}$ is the flux at the turnover frequency.
The values of $\nu_\mathrm{br}$ and $I_\mathrm{0}$ are taken from Fig.~\ref{fig:turnover}.
For the jet width, we use the fit performed in \citet{Boccardi2021}, so that we are able to explore the evolution of the magnetic field as a function of the distance from the core, $B(\mathrm{z})$.
We calculate the magnetic field for each set of parameters shown in Fig.\ \ref{fig:turnover_evolution}.
By performing power-law fits on the obtained $B(\mathrm{z})$ profiles, we find magnetic field evolution indexes in the range between $\sim -0.30$ and $\sim 0.30$.
Consequently, this approach leads to either a very weakly evolving or even increasing magnetic field with the distance from the core.
A magnetic field increasing with the distance from the core is clearly in disagreement with the other methods shown and with the expectations for magnetization in relativistic jets.
While this inconsistency may arise from problems when determining the turnover frequency and flux, for example, due to an inappropriate choice of the frequency range and/or the beam used, we highlight how Eq.\,\ref{eq:B_field_turnover} might be incomplete for the analysis here performed.
Indeed, as shown in \citep{Cawthorne1991, Lobanov1999}, such an equation works properly when exploring the magnetization of a specific jet component, such as a shock, downstream of the jet.
In our case, the goal is to explore the magnetization of the entire bulk flow.
Therefore, we highlight how Eq.\,\ref{eq:B_field_turnover} needs to be adapted for performing such studies, which could prove to be crucial when studying jet magnetization.

\section{Conclusions} \label{sec:Conclusions}

In this paper, we have presented three new VLBA observations of the nearby radio galaxy NGC~315 at 43\,GHz, in addition to other data sets that have been previously reported in the literature.
The main goal of this paper was to investigate the properties of the magnetic field in the inner parsec region by exploring the spectral and brightness temperature properties of the jet.
Our results can be summarized as follows:

\begin{itemize}
    \item We used a pixel-based analysis to extrapolate the jet width in the core region using a 43\,GHz stacked image. The expansion profile suggests the real injection position of the jet to lay $(41\pm 41) \, \mathrm{R_s}$ upstream of the core at 43\,GHz, that is, toward the counter-jet. We find a minimum width of $(35 \pm 4)\, \mathrm{R_s}$, in agreement with the jet width measured in M~87 \citep{Lu2023}. By extrapolating the core-shift function to infinite frequencies, we find slightly larger values for the jet origin position, shifting it more toward the counter-jet. Overall, both methods suggest that the black hole lies approximately $\sim 100 \, \mathrm{R_S}$ upstream of the mm-core.
    \item The spectral index analysis reveals a very steep spectrum on sub-parsec scales, along the acceleration and collimation region.
    Namely, from the 22--43\,GHz spectral index maps at different epochs, we observe values down to $\sim -2$, as also observed in M\,87 \citep{Ro2023}.
    Downstream the jet, on parsec scales, the spectral index reaches more standard values for optically thin emission of about $-0.8$.
    We explored possible physical reasons behind the observed evolution of the spectrum, including cooling due to synchrotron losses to explain the steep values or different acceleration mechanisms between the sub-parsec and parsec region, mainly DSA and magnetic reconnection.
    The DSA is slightly favored for explaining the particle re-acceleration occurring in the jet on all scales. 
    \item We examined the brightness temperature as a function of distance from the core. We infer a relatively flat power-law index at the frequencies that sample the collimation region, or part of it, 
    (22, 43, and 86\,GHz, $-2.9 \lesssim \epsilon\lesssim -2.2$), and a steeper trend ($-3.9 \lesssim \epsilon\lesssim -2.9$) at lower frequencies.
    Moreover, the brightness temperature of the 43\,GHz cores seem to indicate a dominance of the magnetic energy at the jet base, while the cores at progressively lower frequencies reveal a gradual transition towards equipartition, which is reached towards the end of the collimation region. 
    
    \item We determined a synchrotron turnover frequency in the range $10 \, \mathrm{GHz} \lesssim \alpha_\mathrm{br} \lesssim 35 \, \mathrm{GHz}$ at the location of the 43\,GHz core, which decreases down to $\sim 6 \, \mathrm{GHz}$ within $\sim 0.7 \, \mathrm{pc}$. Additionally, we suggest the need for a new formalism to compute the spatial evolution of the magnetic field starting from the determination of the turnover frequency along the jet.
    \item Finally, we employ our observational constraints, together with theoretical models, to explore the geometry of the magnetic field in the inner jet.
    In the conical region, our results are consistent with a toroidal-dominated magnetic field ($a = -1$). In the parabolic region, we derive a magnetic field exponent $-0.80 \lesssim a \lesssim -0.45$, indicating a predominantly toroidal magnetic field that is being dissipated to accelerate the outflow. However, from such an analysis, the contribution of the poloidal component on sub-parsec scales cannot be completely ruled out. 
\end{itemize}

Overall, by utilizing multi-frequency and multi-epoch VLBI datasets, we have extended previous analyses of the magnetic field in NGC 315, suggesting that its toroidal component predominates at the scales investigated.
NGC\,315, given its vicinity and large black hole mass, is an excellent laboratory to test our theories on jet formation and propagation.
Looking ahead, a new research by \citet{Park2024} investigates the observed limb-brightening, while a joint effort between our two teams will aim at exploring the current discrepancy on the extension of the acceleration and collimation region \citep{Boccardi2021, Park2021}.

\bibliographystyle{aa.bst}
\bibliography{bibliography}

\begin{thebibliography}{68}
\expandafter\ifx\csname natexlab\endcsname\relax\def\natexlab#1{#1}\fi

\bibitem[{{Asada} \& {Nakamura}(2012)}]{Asada2012}
{Asada}, K. \& {Nakamura}, M. 2012, \apjl, 745, L28

\bibitem[{{Astropy Collaboration} {et~al.}(2022){Astropy Collaboration},
  {Price-Whelan}, {Lim}, {Earl}, {Starkman}, {Bradley}, {Shupe}, {Patil},
  {Corrales}, {Brasseur}, {N{"o}the}, {Donath}, {Tollerud}, {Morris},
  {Ginsburg}, {Vaher}, {Weaver}, {Tocknell}, {Jamieson}, {van Kerkwijk},
  {Robitaille}, {Merry}, {Bachetti}, {G{"u}nther}, {Aldcroft},
  {Alvarado-Montes}, {Archibald}, {B{'o}di}, {Bapat}, {Barentsen}, {Baz{'a}n},
  {Biswas}, {Boquien}, {Burke}, {Cara}, {Cara}, {Conroy}, {Conseil}, {Craig},
  {Cross}, {Cruz}, {D'Eugenio}, {Dencheva}, {Devillepoix}, {Dietrich},
  {Eigenbrot}, {Erben}, {Ferreira}, {Foreman-Mackey}, {Fox}, {Freij}, {Garg},
  {Geda}, {Glattly}, {Gondhalekar}, {Gordon}, {Grant}, {Greenfield}, {Groener},
  {Guest}, {Gurovich}, {Handberg}, {Hart}, {Hatfield-Dodds}, {Homeier},
  {Hosseinzadeh}, {Jenness}, {Jones}, {Joseph}, {Kalmbach}, {Karamehmetoglu},
  {Ka{l}uszy{'n}ski}, {Kelley}, {Kern}, {Kerzendorf}, {Koch}, {Kulumani},
  {Lee}, {Ly}, {Ma}, {MacBride}, {Maljaars}, {Muna}, {Murphy}, {Norman},
  {O'Steen}, {Oman}, {Pacifici}, {Pascual}, {Pascual-Granado}, {Patil},
  {Perren}, {Pickering}, {Rastogi}, {Roulston}, {Ryan}, {Rykoff}, {Sabater},
  {Sakurikar}, {Salgado}, {Sanghi}, {Saunders}, {Savchenko}, {Schwardt},
  {Seifert-Eckert}, {Shih}, {Jain}, {Shukla}, {Sick}, {Simpson},
  {Singanamalla}, {Singer}, {Singhal}, {Sinha}, {Sip{H{o}}cz}, {Spitler},
  {Stansby}, {Streicher}, {{{S}}umak}, {Swinbank}, {Taranu}, {Tewary},
  {Tremblay}, {Val-Borro}, {Van Kooten}, {Vasovi{'c}}, {Verma}, {de Miranda
  Cardoso}, {Williams}, {Wilson}, {Winkel}, {Wood-Vasey}, {Xue}, {Yoachim},
  {Zhang}, {Zonca}, \& {Astropy Project Contributors}}]{astropy:2022}
{Astropy Collaboration}, {Price-Whelan}, A.~M., {Lim}, P.~L., {et~al.} 2022,
  \apj, 935, 167

\bibitem[{{Astropy Collaboration} {et~al.}(2018){Astropy Collaboration},
  {Price-Whelan}, {Sip{\H{o}}cz}, {G{\"u}nther}, {Lim}, {Crawford}, {Conseil},
  {Shupe}, {Craig}, {Dencheva}, {Ginsburg}, {Vand erPlas}, {Bradley},
  {P{\'e}rez-Su{\'a}rez}, {de Val-Borro}, {Aldcroft}, {Cruz}, {Robitaille},
  {Tollerud}, {Ardelean}, {Babej}, {Bach}, {Bachetti}, {Bakanov}, {Bamford},
  {Barentsen}, {Barmby}, {Baumbach}, {Berry}, {Biscani}, {Boquien}, {Bostroem},
  {Bouma}, {Brammer}, {Bray}, {Breytenbach}, {Buddelmeijer}, {Burke},
  {Calderone}, {Cano Rodr{\'\i}guez}, {Cara}, {Cardoso}, {Cheedella}, {Copin},
  {Corrales}, {Crichton}, {D'Avella}, {Deil}, {Depagne}, {Dietrich}, {Donath},
  {Droettboom}, {Earl}, {Erben}, {Fabbro}, {Ferreira}, {Finethy}, {Fox},
  {Garrison}, {Gibbons}, {Goldstein}, {Gommers}, {Greco}, {Greenfield},
  {Groener}, {Grollier}, {Hagen}, {Hirst}, {Homeier}, {Horton}, {Hosseinzadeh},
  {Hu}, {Hunkeler}, {Ivezi{\'c}}, {Jain}, {Jenness}, {Kanarek}, {Kendrew},
  {Kern}, {Kerzendorf}, {Khvalko}, {King}, {Kirkby}, {Kulkarni}, {Kumar},
  {Lee}, {Lenz}, {Littlefair}, {Ma}, {Macleod}, {Mastropietro}, {McCully},
  {Montagnac}, {Morris}, {Mueller}, {Mumford}, {Muna}, {Murphy}, {Nelson},
  {Nguyen}, {Ninan}, {N{\"o}the}, {Ogaz}, {Oh}, {Parejko}, {Parley}, {Pascual},
  {Patil}, {Patil}, {Plunkett}, {Prochaska}, {Rastogi}, {Reddy Janga},
  {Sabater}, {Sakurikar}, {Seifert}, {Sherbert}, {Sherwood-Taylor}, {Shih},
  {Sick}, {Silbiger}, {Singanamalla}, {Singer}, {Sladen}, {Sooley},
  {Sornarajah}, {Streicher}, {Teuben}, {Thomas}, {Tremblay}, {Turner},
  {Terr{\'o}n}, {van Kerkwijk}, {de la Vega}, {Watkins}, {Weaver}, {Whitmore},
  {Woillez}, {Zabalza}, \& {Astropy Contributors}}]{astropy:2018}
{Astropy Collaboration}, {Price-Whelan}, A.~M., {Sip{\H{o}}cz}, B.~M., {et~al.}
  2018, \aj, 156, 123

\bibitem[{{Astropy Collaboration} {et~al.}(2013){Astropy Collaboration},
  {Robitaille}, {Tollerud}, {Greenfield}, {Droettboom}, {Bray}, {Aldcroft},
  {Davis}, {Ginsburg}, {Price-Whelan}, {Kerzendorf}, {Conley}, {Crighton},
  {Barbary}, {Muna}, {Ferguson}, {Grollier}, {Parikh}, {Nair}, {Unther},
  {Deil}, {Woillez}, {Conseil}, {Kramer}, {Turner}, {Singer}, {Fox}, {Weaver},
  {Zabalza}, {Edwards}, {Azalee Bostroem}, {Burke}, {Casey}, {Crawford},
  {Dencheva}, {Ely}, {Jenness}, {Labrie}, {Lim}, {Pierfederici}, {Pontzen},
  {Ptak}, {Refsdal}, {Servillat}, \& {Streicher}}]{astropy:2013}
{Astropy Collaboration}, {Robitaille}, T.~P., {Tollerud}, E.~J., {et~al.} 2013,
  \aap, 558, A33

\bibitem[{{Baczko} {et~al.}(2022){Baczko}, {Ros}, {Kadler}, {Fromm},
  {Boccardi}, {Perucho}, {Krichbaum}, {Burd}, \& {Zensus}}]{Baczko2022}
{Baczko}, A.~K., {Ros}, E., {Kadler}, M., {et~al.} 2022, \aap, 658, A119

\bibitem[{{Baczko} {et~al.}(2016){Baczko}, {Schulz}, {Kadler}, {Ros},
  {Perucho}, {Krichbaum}, {B{\"o}ck}, {Bremer}, {Grossberger}, {Lindqvist},
  {Lobanov}, {Mannheim}, {Mart{\'\i}-Vidal}, {M{\"u}ller}, {Wilms}, \&
  {Zensus}}]{Baczko2016}
{Baczko}, A.~K., {Schulz}, R., {Kadler}, M., {et~al.} 2016, \aap, 593, A47

\bibitem[{{Blandford} \& {Eichler}(1987)}]{Blandford1987}
{Blandford}, R. \& {Eichler}, D. 1987, \physrep, 154, 1

\bibitem[{{Blandford} {et~al.}(2019){Blandford}, {Meier}, \&
  {Readhead}}]{Blandford2019}
{Blandford}, R., {Meier}, D., \& {Readhead}, A. 2019, \araa, 57, 467

\bibitem[{{Blandford} \& {K{\"o}nigl}(1979)}]{Blandford1979}
{Blandford}, R.~D. \& {K{\"o}nigl}, A. 1979, \apj, 232, 34

\bibitem[{{Blandford} \& {Znajek}(1977)}]{Blandford1977}
{Blandford}, R.~D. \& {Znajek}, R.~L. 1977, \mnras, 179, 433

\bibitem[{{Boccardi} {et~al.}(2021){Boccardi}, {Perucho}, {Casadio}, {Grandi},
  {Macconi}, {Torresi}, {Pellegrini}, {Krichbaum}, {Kadler}, {Giovannini},
  {Karamanavis}, {Ricci}, {Madika}, {Bach}, {Ros}, {Giroletti}, \&
  {Zensus}}]{Boccardi2021}
{Boccardi}, B., {Perucho}, M., {Casadio}, C., {et~al.} 2021, \aap, 647, A67

\bibitem[{{Boizelle} {et~al.}(2021){Boizelle}, {Walsh}, {Barth}, {Buote},
  {Baker}, {Darling}, {Ho}, {Cohn}, \& {Kabasares}}]{Boizelle2021}
{Boizelle}, B.~D., {Walsh}, J.~L., {Barth}, A.~J., {et~al.} 2021, \apj, 908, 19

\bibitem[{{Bower} {et~al.}(2017){Bower}, {Dexter}, {Markoff}, {Rao}, \&
  {Plambeck}}]{Bower2017}
{Bower}, G.~C., {Dexter}, J., {Markoff}, S., {Rao}, R., \& {Plambeck}, R.~L.
  2017, \apjl, 843, L31

\bibitem[{{Cawthorne}(1991)}]{Cawthorne1991}
{Cawthorne}, T.~V. 1991, in Beams and Jets in Astrophysics, Vol.~19, 187

\bibitem[{{Fanaroff} \& {Riley}(1974)}]{FaranoffRiley}
{Fanaroff}, B.~L. \& {Riley}, J.~M. 1974, \mnras, 167, 31P

\bibitem[{{Fromm} {et~al.}(2015){Fromm}, {Fuhrmann}, \& {Perucho}}]{Fromm2015}
{Fromm}, C.~M., {Fuhrmann}, L., \& {Perucho}, M. 2015, \aap, 580, A94

\bibitem[{{Fromm} {et~al.}(2013){Fromm}, {Ros}, {Perucho}, {Savolainen},
  {Mimica}, {Kadler}, {Lobanov}, \& {Zensus}}]{Fromm2013b}
{Fromm}, C.~M., {Ros}, E., {Perucho}, M., {et~al.} 2013, \aap, 557, A105

\bibitem[{{Fuentes} {et~al.}(2018){Fuentes}, {G{\'o}mez}, {Mart{\'\i}}, \&
  {Perucho}}]{Fuentes2018}
{Fuentes}, A., {G{\'o}mez}, J.~L., {Mart{\'\i}}, J.~M., \& {Perucho}, M. 2018,
  \apj, 860, 121

\bibitem[{{Giovannini} {et~al.}(2001){Giovannini}, {Cotton}, {Feretti}, {Lara},
  \& {Venturi}}]{Giovannini2001}
{Giovannini}, G., {Cotton}, W.~D., {Feretti}, L., {Lara}, L., \& {Venturi}, T.
  2001, \apj, 552, 508

\bibitem[{{Greisen}(1990)}]{Greisen1990}
{Greisen}, E.~W. 1990, in Acquisition, Processing and Archiving of Astronomical
  Images, 125--142

\bibitem[{{Hada} {et~al.}(2011){Hada}, {Doi}, {Kino}, {Nagai}, {Hagiwara}, \&
  {Kawaguchi}}]{Hada2011}
{Hada}, K., {Doi}, A., {Kino}, M., {et~al.} 2011, \nat, 477, 185

\bibitem[{{Hovatta} {et~al.}(2014){Hovatta}, {Aller}, {Aller}, {Clausen-Brown},
  {Homan}, {Kovalev}, {Lister}, {Pushkarev}, \& {Savolainen}}]{Hovatta2014}
{Hovatta}, T., {Aller}, M.~F., {Aller}, H.~D., {et~al.} 2014, \aj, 147, 143

\bibitem[{{Kadler} {et~al.}(2004){Kadler}, {Ros}, {Lobanov}, {Falcke}, \&
  {Zensus}}]{Kadler2004}
{Kadler}, M., {Ros}, E., {Lobanov}, A.~P., {Falcke}, H., \& {Zensus}, J.~A.
  2004, \aap, 426, 481

\bibitem[{{Kardashev}(1962)}]{Kardashev1962}
{Kardashev}, N.~S. 1962, \sovast, 6, 317

\bibitem[{{Kellermann} \& {Pauliny-Toth}(1969)}]{Kellermann1969}
{Kellermann}, K.~I. \& {Pauliny-Toth}, I.~I.~K. 1969, \apjl, 155, L71

\bibitem[{{Kino} {et~al.}(2024){Kino}, {Ro}, {Takahashi}, {Kawashima}, {Park},
  {Hada}, \& {Cui}}]{Kino2024}
{Kino}, M., {Ro}, H., {Takahashi}, M., {et~al.} 2024, \apj, 973, 100

\bibitem[{{Komatsu} {et~al.}(2009){Komatsu}, {Dunkley}, {Nolta}, {Bennett},
  {Gold}, {Hinshaw}, {Jarosik}, {Larson}, {Limon}, {Page}, {Spergel},
  {Halpern}, {Hill}, {Kogut}, {Meyer}, {Tucker}, {Weiland}, {Wollack}, \&
  {Wright}}]{Komatsu}
{Komatsu}, E., {Dunkley}, J., {Nolta}, M.~R., {et~al.} 2009, \apjs, 180, 330

\bibitem[{{Komissarov}(2012)}]{Komissarov2012}
{Komissarov}, S.~S. 2012, \mnras, 422, 326

\bibitem[{{Komissarov} {et~al.}(2007){Komissarov}, {Barkov}, {Vlahakis}, \&
  {K{\"o}nigl}}]{Komissarov2007}
{Komissarov}, S.~S., {Barkov}, M.~V., {Vlahakis}, N., \& {K{\"o}nigl}, A. 2007,
  \mnras, 380, 51

\bibitem[{{Kovalev} {et~al.}(2020){Kovalev}, {Pushkarev}, {Nokhrina}, {Plavin},
  {Beskin}, {Chernoglazov}, {Lister}, \& {Savolainen}}]{Kovalev2020}
{Kovalev}, Y.~Y., {Pushkarev}, A.~B., {Nokhrina}, E.~E., {et~al.} 2020, \mnras,
  495, 3576

\bibitem[{{L{\"a}hteenm{\"a}ki} \& {Valtaoja}(1999)}]{Lahte1999}
{L{\"a}hteenm{\"a}ki}, A. \& {Valtaoja}, E. 1999, \apj, 521, 493

\bibitem[{{Laing} {et~al.}(2006){Laing}, {Canvin}, {Cotton}, \&
  {Bridle}}]{Laing2006}
{Laing}, R.~A., {Canvin}, J.~R., {Cotton}, W.~D., \& {Bridle}, A.~H. 2006,
  \mnras, 368, 48

\bibitem[{{Lister} {et~al.}(2018){Lister}, {Aller}, {Aller}, {Hodge}, {Homan},
  {Kovalev}, {Pushkarev}, \& {Savolainen}}]{Lister2018}
{Lister}, M.~L., {Aller}, M.~F., {Aller}, H.~D., {et~al.} 2018, \apjs, 234, 12

\bibitem[{{Lister} {et~al.}(2019){Lister}, {Homan}, {Hovatta}, {Kellermann},
  {Kiehlmann}, {Kovalev}, {Max-Moerbeck}, {Pushkarev}, {Readhead}, {Ros}, \&
  {Savolainen}}]{Lister2019}
{Lister}, M.~L., {Homan}, D.~C., {Hovatta}, T., {et~al.} 2019, \apj, 874, 43

\bibitem[{{Lobanov}(1998)}]{Lobanov1998}
{Lobanov}, A.~P. 1998, \aaps, 132, 261

\bibitem[{{Lobanov} \& {Zensus}(1999)}]{Lobanov1999}
{Lobanov}, A.~P. \& {Zensus}, J.~A. 1999, \apj, 521, 509

\bibitem[{{Loureiro} \& {Uzdensky}(2016)}]{Loureiro2016}
{Loureiro}, N.~F. \& {Uzdensky}, D.~A. 2016, Plasma Physics and Controlled
  Fusion, 58, 014021

\bibitem[{{Lu} {et~al.}(2023){Lu}, {Asada}, {Krichbaum}, {Park}, {Tazaki},
  {Pu}, {Nakamura}, {Lobanov}, {Hada}, {Akiyama}, {Kim}, {Marti-Vidal},
  {G{\'o}mez}, {Kawashima}, {Yuan}, {Ros}, {Alef}, {Britzen}, {Bremer},
  {Broderick}, {Doi}, {Giovannini}, {Giroletti}, {Ho}, {Honma}, {Hughes},
  {Inoue}, {Jiang}, {Kino}, {Koyama}, {Lindqvist}, {Liu}, {Marscher},
  {Matsushita}, {Nagai}, {Rottmann}, {Savolainen}, {Schuster}, {Shen}, {de
  Vicente}, {Walker}, {Yang}, {Zensus}, {Algaba}, {Allardi}, {Bach},
  {Berthold}, {Bintley}, {Byun}, {Casadio}, {Chang}, {Chang}, {Chang}, {Chen},
  {Chen}, {Chilson}, {Chuter}, {Conway}, {Crew}, {Dempsey}, {Dornbusch},
  {Faber}, {Friberg}, {Garc{\'\i}a}, {Garrido}, {Han}, {Han}, {Hasegawa},
  {Herrero-Illana}, {Huang}, {Huang}, {Impellizzeri}, {Jiang}, {Jinchi},
  {Jung}, {Kallunki}, {Kirves}, {Kimura}, {Koay}, {Koch}, {Kramer}, {Kraus},
  {Kubo}, {Kuo}, {Li}, {Lin}, {Liu}, {Liu}, {Lo}, {Lu}, {MacDonald},
  {Martin-Cocher}, {Messias}, {Meyer-Zhao}, {Minter}, {Nair}, {Nishioka},
  {Norton}, {Nystrom}, {Ogawa}, {Oshiro}, {Patel}, {Pen}, {Pidopryhora},
  {Pradel}, {Raffin}, {Rao}, {Ruiz}, {Sanchez}, {Shaw}, {Snow}, {Sridharan},
  {Srinivasan}, {Tercero}, {Torne}, {Traianou}, {Wagner}, {Walther}, {Wei},
  {Yang}, \& {Yu}}]{Lu2023}
{Lu}, R.-S., {Asada}, K., {Krichbaum}, T.~P., {et~al.} 2023, \nat, 616, 686

\bibitem[{{Marscher} \& {Gear}(1985)}]{Marscher1985}
{Marscher}, A.~P. \& {Gear}, W.~K. 1985, \apj, 298, 114

\bibitem[{{Marscher} {et~al.}(2008){Marscher}, {Jorstad}, {D'Arcangelo},
  {Smith}, {Williams}, {Larionov}, {Oh}, {Olmstead}, {Aller}, {Aller},
  {McHardy}, {L{\"a}hteenm{\"a}ki}, {Tornikoski}, {Valtaoja}, {Hagen-Thorn},
  {Kopatskaya}, {Gear}, {Tosti}, {Kurtanidze}, {Nikolashvili}, {Sigua},
  {Miller}, \& {Ryle}}]{Marscher2008}
{Marscher}, A.~P., {Jorstad}, S.~G., {D'Arcangelo}, F.~D., {et~al.} 2008, \nat,
  452, 966

\bibitem[{{Meenakshi} {et~al.}(2023){Meenakshi}, {Mukherjee}, {Bodo}, \&
  {Rossi}}]{Meenakhsi2023}
{Meenakshi}, M., {Mukherjee}, D., {Bodo}, G., \& {Rossi}, P. 2023, \mnras, 526,
  5418

\bibitem[{{Nikonov} {et~al.}(2023){Nikonov}, {Kovalev}, {Kravchenko},
  {Pashchenko}, \& {Lobanov}}]{Nikonov2023}
{Nikonov}, A.~S., {Kovalev}, Y.~Y., {Kravchenko}, E.~V., {Pashchenko}, I.~N.,
  \& {Lobanov}, A.~P. 2023, arXiv e-prints, arXiv:2307.11660

\bibitem[{{O'Sullivan} \& {Gabuzda}(2009)}]{OSullivan2009}
{O'Sullivan}, S.~P. \& {Gabuzda}, D.~C. 2009, \mnras, 400, 26

\bibitem[{{Pacholczyk}(1970)}]{Pacho1970}
{Pacholczyk}, A.~G. 1970, {Radio astrophysics. Nonthermal processes in galactic
  and extragalactic sources}

\bibitem[{{Park} {et~al.}(2021{\natexlab{a}}){Park}, {Byun}, {Asada}, \&
  {Yun}}]{Park2021_GPCAL}
{Park}, J., {Byun}, D.-Y., {Asada}, K., \& {Yun}, Y. 2021{\natexlab{a}}, \apj,
  906, 85

\bibitem[{{Park} {et~al.}(2021{\natexlab{b}}){Park}, {Hada}, {Nakamura},
  {Asada}, {Zhao}, \& {Kino}}]{Park2021}
{Park}, J., {Hada}, K., {Nakamura}, M., {et~al.} 2021{\natexlab{b}}, \apj, 909,
  76

\bibitem[{{Park} {et~al.}(2024){Park}, {Zhao}, {Nakamura}, {Mizuno}, {Pu},
  {Asada}, {Takahashi}, {Toma}, {Kino}, {Cho}, {Hada}, {Edwards}, {Ro}, {Kam},
  {Yi}, {Lee}, {Koyama}, {Byun}, {Phillips}, {Reynolds}, {Hodgson}, \&
  {Lee}}]{Park2024}
{Park}, J., {Zhao}, G.-Y., {Nakamura}, M., {et~al.} 2024, \apjl, 973, L45

\bibitem[{{Pushkarev} {et~al.}(2017){Pushkarev}, {Kovalev}, {Lister},
  {Savolainen}, {Aller}, {Aller}, \& {Hodge}}]{Pushkarev2017b}
{Pushkarev}, A.~B., {Kovalev}, Y.~Y., {Lister}, M.~L., {et~al.} 2017, Galaxies,
  5, 93

\bibitem[{{Readhead}(1994)}]{Readhead1994}
{Readhead}, A. C.~S. 1994, \apj, 426, 51

\bibitem[{{Ricci} {et~al.}(2022){Ricci}, {Boccardi}, {Nokhrina}, {Perucho},
  {MacDonald}, {Mattia}, {Grandi}, {Madika}, {Krichbaum}, \&
  {Zensus}}]{Ricci2022}
{Ricci}, L., {Boccardi}, B., {Nokhrina}, E., {et~al.} 2022, \aap, 664, A166

\bibitem[{{Ricci} {et~al.}(2024){Ricci}, {Perucho}, {L{\'o}pez-Miralles},
  {Mart{\'\i}}, \& {Boccardi}}]{Ricci2024}
{Ricci}, L., {Perucho}, M., {L{\'o}pez-Miralles}, J., {Mart{\'\i}}, J.~M., \&
  {Boccardi}, B. 2024, \aap, 683, A235

\bibitem[{{Rieger} \& {Duffy}(2019)}]{Rieger2019}
{Rieger}, F.~M. \& {Duffy}, P. 2019, \apjl, 886, L26

\bibitem[{{Rieger} \& {Levinson}(2018)}]{Rieger2018}
{Rieger}, F.~M. \& {Levinson}, A. 2018, Galaxies, 6, 116

\bibitem[{{Ripperda} {et~al.}(2019){Ripperda}, {Porth}, {Sironi}, \&
  {Keppens}}]{Ripperda2019}
{Ripperda}, B., {Porth}, O., {Sironi}, L., \& {Keppens}, R. 2019, \mnras, 485,
  299

\bibitem[{{Ro} {et~al.}(2023){Ro}, {Kino}, {Sohn}, {Hada}, {Park}, {Nakamura},
  {Cui}, {Yi}, {Chung}, {Hodgson}, {Kawashima}, {An}, {Trippe}, {Algaba},
  {Kim}, {Sawada-Satoh}, {Wajima}, {Shen}, {Cheng}, {Cho}, {Jiang}, {Jung},
  {Lee}, {Niinuma}, {Oh}, {Tazaki}, {Zhao}, {Akiyama}, {Honma}, {Lee}, {Lu},
  {Zhang}, {Asada}, {Cui}, {Hagiwara}, {Hirota}, {Kawaguchi}, {Koyama}, {Lee},
  {Oh}, {Sugiyama}, {Takamura}, {Wang}, {Hwang}, {Jung}, {Kim}, {Kim},
  {Kobayashi}, {Oh}, {Oyama}, {Roh}, \& {Yeom}}]{Ro2023}
{Ro}, H., {Kino}, M., {Sohn}, B.~W., {et~al.} 2023, arXiv e-prints,
  arXiv:2303.01014

\bibitem[{{Schinzel} {et~al.}(2012){Schinzel}, {Lobanov}, {Taylor}, {Jorstad},
  {Marscher}, \& {Zensus}}]{Schinzel2012}
{Schinzel}, F.~K., {Lobanov}, A.~P., {Taylor}, G.~B., {et~al.} 2012, \aap, 537,
  A70

\bibitem[{{Seo} {et~al.}(2023){Seo}, {Ryu}, \& {Kang}}]{Seo2023}
{Seo}, J., {Ryu}, D., \& {Kang}, H. 2023, \apj, 944, 199

\bibitem[{{Shepherd}(1997)}]{Shepherd1997}
{Shepherd}, M.~C. 1997, in Astronomical Society of the Pacific Conference
  Series, Vol. 125, Astronomical Data Analysis Software and Systems VI, ed.
  G.~{Hunt} \& H.~{Payne}, 77

\bibitem[{{Singal}(2009)}]{Singal2009}
{Singal}, A.~K. 2009, \apjl, 703, L109

\bibitem[{{Sironi} {et~al.}(2016){Sironi}, {Giannios}, \&
  {Petropoulou}}]{Sironi2016}
{Sironi}, L., {Giannios}, D., \& {Petropoulou}, M. 2016, \mnras, 462, 48

\bibitem[{{Sironi} \& {Spitkovsky}(2009)}]{Sironi2009}
{Sironi}, L. \& {Spitkovsky}, A. 2009, \apj, 698, 1523

\bibitem[{{Sironi} \& {Spitkovsky}(2011)}]{Sironi2011}
{Sironi}, L. \& {Spitkovsky}, A. 2011, \apj, 726, 75

\bibitem[{{Sironi} \& {Spitkovsky}(2014)}]{Sironi2014}
{Sironi}, L. \& {Spitkovsky}, A. 2014, \apjl, 783, L21

\bibitem[{{Sokoloff} {et~al.}(1998){Sokoloff}, {Bykov}, {Shukurov},
  {Berkhuijsen}, {Beck}, \& {Poezd}}]{Sokoloff1998}
{Sokoloff}, D.~D., {Bykov}, A.~A., {Shukurov}, A., {et~al.} 1998, \mnras, 299,
  189

\bibitem[{{Trager} {et~al.}(2000){Trager}, {Faber}, {Worthey}, \&
  {Gonz{\'a}lez}}]{Trager2000}
{Trager}, S.~C., {Faber}, S.~M., {Worthey}, G., \& {Gonz{\'a}lez}, J.~J. 2000,
  \aj, 119, 1645

\bibitem[{{Vlahakis} \& {K{\"o}nigl}(2003{\natexlab{a}})}]{Vlahakis2003_a}
{Vlahakis}, N. \& {K{\"o}nigl}, A. 2003{\natexlab{a}}, \apj, 596, 1080

\bibitem[{{Vlahakis} \& {K{\"o}nigl}(2003{\natexlab{b}})}]{Vlahakis2003_b}
{Vlahakis}, N. \& {K{\"o}nigl}, A. 2003{\natexlab{b}}, \apj, 596, 1104

\bibitem[{{Vlahakis} \& {K{\"o}nigl}(2004)}]{Vlahakis2004}
{Vlahakis}, N. \& {K{\"o}nigl}, A. 2004, \apj, 605, 656

\end{thebibliography}

\begin{acknowledgements} 
The authors would like to thank the referee for the insightful comments which improved the manuscript.
The authors thank Christian Fromm for its support in discussions on the results.
LR is funded by the Deutsche Forschungsgemeinschaft (DFG, German Research Foundation) – project number 443220636. 
LR, BB, and EM acknowledge the financial support of a Otto Hahn research group from the Max Planck Society.
MP acknowledges support from \texttt{MICIU/AEI/10.13039/501100011033} and FEDER, UE, via the grant PID2022-136828NB-C43, from the Generalitat Valenciana through grant CIPROM/2022/49, and from the Astrophysics and High Energy Physics project programme supported by the Spanish Ministry of Science and Generalitat Valenciana with funding from European Union NextGenerationEU (\texttt{PRTR-C17.I1}) through grant \texttt{ASFAE/2022/005}. JR and PB received financial support for this research from the International Max Planck Research School (IMPRS) for Astronomy and Astrophysics at the Universities of Bonn and Cologne.
This research has made use of data from the MOJAVE database that is maintained by the MOJAVE team \citep{Lister2018}. 
The VLBA is a facility of the National Science Foundation under cooperative agreement by Associated Universities, Inc.
This work made use of Astropy:\footnote{http://www.astropy.org} a community-developed core Python package and an ecosystem of tools and resources for astronomy \citep{astropy:2013, astropy:2018, astropy:2022}.
\end{acknowledgements}

\begin{appendix}

\section{Flux density} \label{app:data}

\begin{figure}[t]
    \centering
    \includegraphics[width=\linewidth]{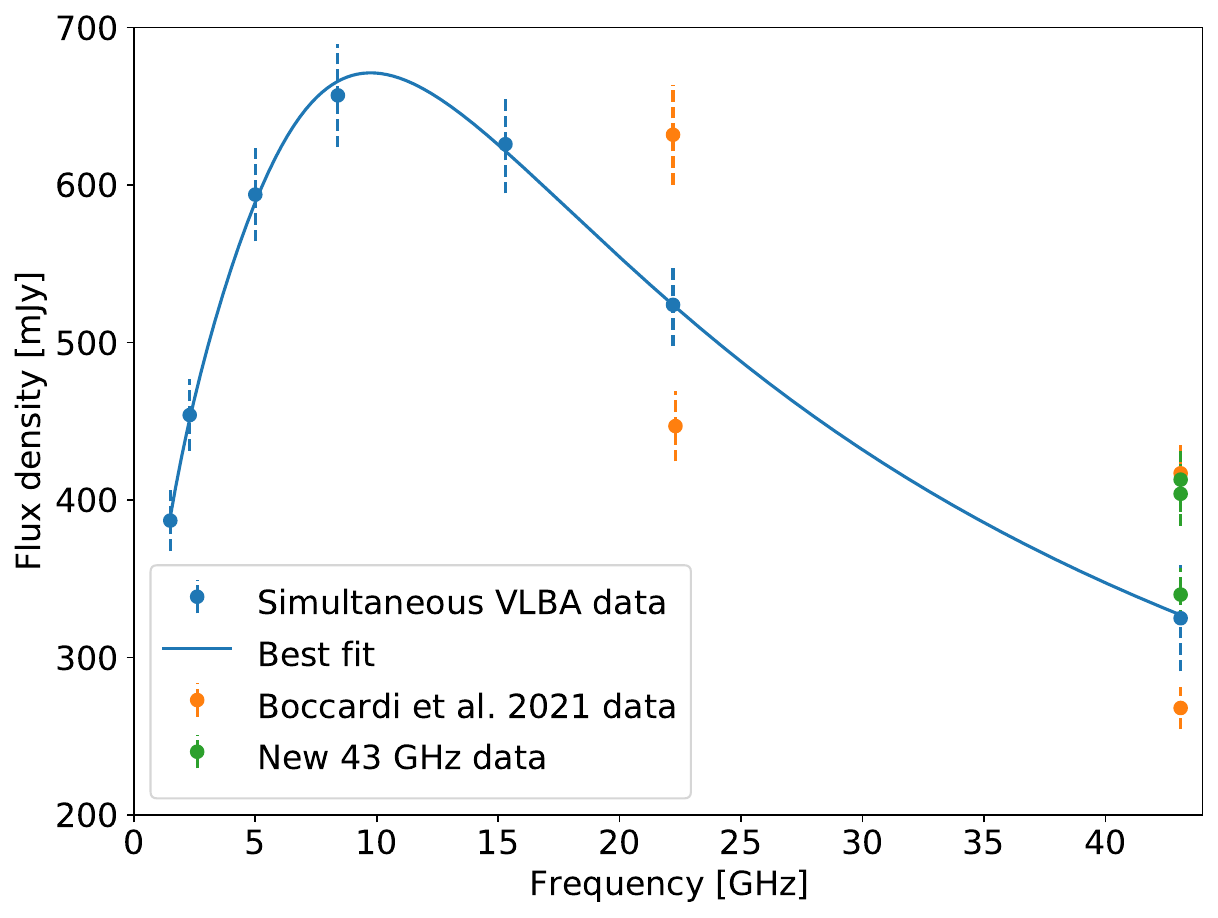}\par
    \caption{Flux density as a function of frequency. The blue data points are from the re-analyzed multi-frequency VLBA data set, the orange points are from the 22\,GHz and 43\,GHz observations published in \citet{Boccardi2021}, and the green points are from the new 43\,GHz observations presented in Sect.~\ref{sec:original_data}. 
    }
    \label{fig:fluxes}
\end{figure}

In Fig.~\ref{fig:fluxes} we show the flux density as a function of the observing frequency for the maps used in the spectral analysis, along with the new 43\,GHz observations.
We fix the errors on the flux density at 5\%, except for the 43\,GHz map from 2020, where we assume a more conservative error of 10\% considering the low data quality.

In order to verify the consistency of the total flux densities obtained for the simultaneous VLBA data set, we fit a synchrotron emission spectrum to the data \citep{Pacho1970}:
\begin{equation}
    I_\nu = I_0 \left( \frac{\nu}{\nu_\mathrm{br}} \right)^{\alpha_t} \left[ 1 - \mathrm{exp} \left( - \left( \frac{\nu_\mathrm{br}}{\nu} \right)^{\alpha_t - \alpha} \right)  \right] \, .
    \label{eq:syn_turn}
\end{equation}
Here, $I_0$ is the flux density at the turnover, $\nu_\mathrm{br}$ is the turnover frequency, $\alpha_t$ is the spectral index of the optically thick region, and $\alpha$ of the optically thin one.
All the parameters are left to vary freely.
Shown as a blue line in Fig.~\ref{fig:fluxes}, the flux density dependence on the frequency is well described by pure synchrotron emission.
A comparison with data obtained at different epochs at 22\,GHz and 43\,GHz shows, however, a significant scatter, indicating that NGC~315 is quite variable at high radio frequencies.

\section{Gaussian components} \label{app:gaussian_comp}

\begin{table}[h]
\centering
\caption{Modelfit parameters of the Gaussian components at 1\,GHz at the different epochs.} 
\begin{tabular}{c|ccccc}
\hline

Obs date & \begin{tabular}[c]{@{}c@{}}S\\ {[}mJy{]}\end{tabular} & \begin{tabular}[c]{@{}c@{}}d\\ {[}mas{]}\end{tabular} & \begin{tabular}[c]{@{}c@{}}z\\ {[}mas{]}\end{tabular} & \begin{tabular}[c]{@{}c@{}}PA\\ {[}deg{]}\end{tabular} \\ \hline

Jan 2020  & 2.68 & 7.96 & -6.54 & 44.98 \\
 & 80.24 & 0.5 & 2.37 & -47.0 \\
 & 150.62 & 0.2 & 5.13 & -49.51 \\
 & 66.92 & 0.69 & 8.71 & -49.89 \\
 & 26.22 & 0.93 & 13.0 & -49.27 \\
 & 11.82 & 2.08 & 17.4 & -49.62 \\
 & 11.93 & 2.6 & 24.34 & -50.51 \\
 & 8.6 & 2.72 & 29.43 & -50.58 \\
 & 4.47 & 4.19 & 36.91 & -50.66 \\
 & 5.53 & 3.44 & 44.13 & -50.91 \\
 & 4.08 & 5.77 & 51.97 & -49.46 \\
 & 6.89 & 35.4 & 79.77 & -50.77 \\

\hline

\end{tabular}
\begin{flushleft}
\textbf{Notes.} Column 1: date of the observation; Column 2: flux density in Jy; Column 3: FWHM in mas; Column 4: radial distance after the core-shift correction; Column 5: position angle after the core-shift correction. 
\end{flushleft}
\label{tab:table_modelfit}
\end{table}

\begin{table}[h]
\centering
\caption{Modelfit parameters of the Gaussian components at 2\,GHz at the different epochs.} 
\centering
\begin{tabular}{c|ccccc}
\hline

Obs date & \begin{tabular}[c]{@{}c@{}}S\\ {[}mJy{]}\end{tabular} & \begin{tabular}[c]{@{}c@{}}d\\ {[}mas{]}\end{tabular} & \begin{tabular}[c]{@{}c@{}}z\\ {[}mas{]}\end{tabular} & \begin{tabular}[c]{@{}c@{}}PA\\ {[}deg{]}\end{tabular} \\ \hline
Jan 2020  & 6.05 & 4.6 & -5.18 & 30.29 \\
 & 199.94 & 0.64 & 2.93 & -52.89 \\
 & 144.74 & 0.93 & 5.75 & -51.79 \\
 & 46.14 & 1.33 & 9.79 & -49.43 \\
 & 27.66 & 3.15 & 14.27 & -50.83 \\
 & 15.92 & 3.77 & 26.41 & -51.32 \\
 & 12.53 & 14.92 & 43.63 & -51.49 \\

\hline

\end{tabular}
\begin{flushleft}
\textbf{Notes.} See Table \ref{tab:table_modelfit}
\end{flushleft}
\end{table}

\begin{table}[h]
\centering
\caption{Modelfit parameters of the Gaussian components at 5\,GHz at the different epochs.} 
\begin{tabular}{c|ccccc}
\hline

Obs date & \begin{tabular}[c]{@{}c@{}}S\\ {[}mJy{]}\end{tabular} & \begin{tabular}[c]{@{}c@{}}d\\ {[}mas{]}\end{tabular} & \begin{tabular}[c]{@{}c@{}}z\\ {[}mas{]}\end{tabular} & \begin{tabular}[c]{@{}c@{}}PA\\ {[}deg{]}\end{tabular} \\ \hline

Jan 2020  & 6.14 & 0.07 & -0.59 & 53.52 \\
 & 211.22 & 0.07 & 0.89 & -45.0 \\
 & 188.45 & 0.33 & 1.83 & -46.07 \\
 & 63.07 & 0.57 & 3.48 & -48.2 \\
 & 58.75 & 0.9 & 5.26 & -49.7 \\
 & 3.6 & 0.1 & 6.58 & -48.87 \\
 & 29.37 & 1.7 & 7.95 & -49.5 \\
 & 3.88 & 0.08 & 9.74 & -50.5 \\
 & 17.06 & 2.91 & 11.94 & -49.51 \\
 & 3.67 & 1.26 & 16.03 & -48.15 \\
 & 1.09 & 3.01 & 20.09 & -50.52 \\
 & 2.6 & 2.77 & 25.05 & -50.32 \\
 & 1.45 & 2.38 & 29.5 & -50.96 \\

\hline

\end{tabular}
\begin{flushleft}
\textbf{Notes.} See Table \ref{tab:table_modelfit}
\end{flushleft}
\end{table}

\begin{table}[]
\centering
\caption{Modelfit parameters of the Gaussian components at 8\,GHz at the different epochs.} 
\begin{tabular}{c|ccccc}
\hline

Obs date & \begin{tabular}[c]{@{}c@{}}S\\ {[}mJy{]}\end{tabular} & \begin{tabular}[c]{@{}c@{}}d\\ {[}mas{]}\end{tabular} & \begin{tabular}[c]{@{}c@{}}z\\ {[}mas{]}\end{tabular} & \begin{tabular}[c]{@{}c@{}}PA\\ {[}deg{]}\end{tabular} \\ \hline

Jan 2020  & 2.42 & 2.0 & -3.06 & 44.11 \\
 & 82.37 & 0.08 & -0.05 & 102.05 \\
 & 264.47 & 0.1 & 0.57 & -45.0 \\
 & 139.98 & 0.15 & 1.21 & -45.16 \\
 & 47.5 & 0.22 & 2.02 & -47.83 \\
 & 22.91 & 0.38 & 3.02 & -47.93 \\
 & 25.94 & 0.53 & 3.95 & -48.73 \\
 & 24.38 & 0.43 & 5.06 & -50.42 \\
 & 10.11 & 0.48 & 6.01 & -49.63 \\
 & 8.89 & 0.73 & 7.46 & -49.25 \\
 & 8.54 & 0.83 & 8.79 & -50.61 \\
 & 1.02 & 0.15 & 9.85 & -50.48 \\
 & 4.92 & 1.4 & 11.24 & -50.0 \\
 & 3.38 & 1.31 & 13.82 & -48.85 \\
 & 2.07 & 1.59 & 16.44 & -49.76 \\
 & 3.08 & 2.55 & 24.93 & -50.87 \\
 
\hline

\end{tabular}
\begin{flushleft}
\textbf{Notes.} See Table \ref{tab:table_modelfit}
\end{flushleft}
\end{table}

\begin{table}[]
\centering
\caption{Modelfit parameters of the Gaussian components at 15\,GHz at the different epochs.} 
\begin{tabular}{c|ccccc}
\hline

Obs date & \begin{tabular}[c]{@{}c@{}}S\\ {[}mJy{]}\end{tabular} & \begin{tabular}[c]{@{}c@{}}d\\ {[}mas{]}\end{tabular} & \begin{tabular}[c]{@{}c@{}}z\\ {[}mas{]}\end{tabular} & \begin{tabular}[c]{@{}c@{}}PA\\ {[}deg{]}\end{tabular} \\ \hline

Apr 1995  & 159.12 & 0.05 & 0.08 & -52.97 \\
 & 292.61 & 0.07 & 0.36 & -56.31 \\
 & 146.87 & 0.17 & 0.77 & -55.43 \\
 & 82.15 & 0.3 & 1.33 & -54.54 \\
 & 74.99 & 0.45 & 1.96 & -53.59 \\
 & 23.05 & 0.59 & 3.5 & -52.44 \\
 & 21.06 & 0.62 & 4.54 & -48.81 \\

\hline

Dec 1995  & 91.51 & 0.05 & 0.04 & -117.9 \\
 & 270.98 & 0.13 & 0.36 & -56.31 \\
 & 119.63 & 0.2 & 0.75 & -55.95 \\
 & 65.95 & 0.17 & 1.22 & -54.16 \\
 & 49.68 & 0.31 & 1.7 & -53.93 \\
 & 37.02 & 0.33 & 2.27 & -53.63 \\
 & 33.39 & 0.89 & 3.89 & -51.45 \\
 & 16.48 & 1.19 & 5.34 & -50.37 \\

\hline

May 1996  & 69.92 & 0.07 & 0.03 & -122.51 \\
 & 255.53 & 0.09 & 0.36 & -56.31 \\
 & 103.15 & 0.14 & 0.78 & -54.23 \\
 & 55.82 & 0.09 & 1.2 & -54.37 \\
 & 48.11 & 0.28 & 1.71 & -53.82 \\
 & 25.34 & 0.32 & 2.34 & -54.2 \\
 & 19.11 & 0.55 & 3.34 & -52.81 \\
 & 28.57 & 0.69 & 4.57 & -50.03 \\

\hline

Jul 1996  & 98.22 & 0.11 & 0.06 & -67.77 \\
 & 194.48 & 0.07 & 0.36 & -56.31 \\
 & 90.52 & 0.12 & 0.74 & -56.73 \\
 & 63.95 & 0.19 & 1.16 & -54.43 \\
 & 41.29 & 0.29 & 1.71 & -53.71 \\
 & 24.6 & 0.33 & 2.23 & -54.17 \\
 & 10.94 & 0.5 & 2.95 & -53.22 \\
 & 12.46 & 0.48 & 3.83 & -52.26 \\
 & 16.37 & 0.62 & 4.65 & -49.7 \\
 & 9.71 & 0.67 & 6.16 & -50.74 \\

\hline

Aug 1997  & 145.67 & 0.14 & 0.0 & -25.97 \\
 & 280.4 & 0.08 & 0.36 & -47.96 \\
 & 136.95 & 0.24 & 0.84 & -52.36 \\
 & 54.57 & 0.2 & 1.41 & -53.86 \\
 & 41.8 & 0.45 & 2.01 & -53.38 \\
 & 18.07 & 0.81 & 2.94 & -54.13 \\
 & 13.62 & 0.4 & 4.04 & -50.96 \\
 & 16.04 & 0.75 & 4.79 & -50.29 \\
 & 10.17 & 0.79 & 6.32 & -50.11 \\
 & 5.21 & 1.01 & 7.7 & -49.47 \\

\hline

Jul 1999  & 178.51 & 0.07 & -0.07 & 18.52 \\
 & 338.76 & 0.13 & 0.36 & -56.31 \\
 & 118.02 & 0.1 & 0.96 & -52.03 \\
 & 51.31 & 0.3 & 1.66 & -52.59 \\
 & 35.06 & 0.22 & 2.4 & -54.14 \\
 & 19.55 & 0.85 & 3.46 & -53.34 \\
 & 9.53 & 0.44 & 4.5 & -49.77 \\
 & 18.6 & 2.04 & 6.54 & -52.38 \\

\hline

\end{tabular}
\begin{flushleft}
\textbf{Notes.} See Table \ref{tab:table_modelfit}
\end{flushleft}
\end{table}

\begin{table}[]
\centering
\caption{Modelfit parameters of the Gaussian components at 15\,GHz at the different epochs.} 
\begin{tabular}{c|ccccc}
\hline

Obs date & \begin{tabular}[c]{@{}c@{}}S\\ {[}mJy{]}\end{tabular} & \begin{tabular}[c]{@{}c@{}}d\\ {[}mas{]}\end{tabular} & \begin{tabular}[c]{@{}c@{}}z\\ {[}mas{]}\end{tabular} & \begin{tabular}[c]{@{}c@{}}PA\\ {[}deg{]}\end{tabular} \\ \hline

Dec 2000  & 6.05 & 0.04 & 0.14 & -177.68 \\
 & 182.41 & 0.15 & 0.36 & -56.31 \\
 & 130.9 & 0.23 & 0.8 & -51.47 \\
 & 78.65 & 0.27 & 1.34 & -52.26 \\
 & 44.84 & 0.33 & 2.02 & -52.33 \\
 & 12.6 & 0.01 & 2.72 & -52.62 \\
 & 12.2 & 0.51 & 3.49 & -53.78 \\
 & 19.79 & 0.68 & 4.61 & -50.98 \\
 & 6.59 & 0.64 & 6.53 & -44.51 \\

\hline

Oct 2003  & 297.83 & 0.17 & 0.36 & -56.31 \\
 & 123.97 & 0.2 & 0.9 & -52.85 \\
 & 75.16 & 0.32 & 1.49 & -50.85 \\
 & 29.02 & 0.25 & 2.16 & -52.45 \\
 & 36.3 & 0.99 & 3.07 & -53.3 \\
 & 4.98 & 0.39 & 4.15 & -55.45 \\
 & 6.29 & 0.49 & 4.94 & -51.04 \\

\hline

Feb 2006  & 86.88 & 0.04 & 0.07 & -156.03 \\
 & 277.98 & 0.01 & 0.36 & -56.31 \\
 & 96.63 & 0.12 & 0.74 & -56.42 \\
 & 60.51 & 0.16 & 1.24 & -52.21 \\
 & 33.13 & 0.35 & 1.81 & -52.1 \\
 & 30.3 & 0.63 & 2.82 & -52.15 \\
 & 21.72 & 0.86 & 4.51 & -51.27 \\
 & 14.24 & 1.19 & 6.28 & -50.61 \\

\hline

Sep 2008 & 246.26 & 0.07 & 0.36 & -56.31 \\
 & 144.46 & 0.09 & 0.7 & -54.5 \\
 & 83.56 & 0.15 & 1.13 & -52.94 \\
 & 50.61 & 0.2 & 1.63 & -52.95 \\
 & 32.65 & 0.31 & 2.16 & -50.81 \\
 & 20.19 & 0.38 & 2.89 & -50.85 \\
 & 8.92 & 0.49 & 3.61 & -53.03 \\
 & 13.75 & 0.66 & 4.76 & -51.67 \\
 & 6.79 & 0.68 & 6.17 & -50.71 \\
 & 8.29 & 0.96 & 7.56 & -51.17 \\
 & 2.27 & 0.79 & 9.11 & -51.84 \\

\hline

Jun 2009  & 51.06 & 0.02 & 0.05 & -129.28 \\
 & 232.51 & 0.09 & 0.36 & -56.31 \\
 & 171.09 & 0.06 & 0.67 & -53.55 \\
 & 73.63 & 0.19 & 1.12 & -52.75 \\
 & 37.26 & 0.21 & 1.62 & -52.55 \\
 & 26.9 & 0.27 & 2.14 & -52.1 \\
 & 16.25 & 0.31 & 2.62 & -51.5 \\
 & 13.96 & 0.42 & 3.29 & -50.59 \\
 & 8.33 & 0.67 & 4.2 & -51.94 \\
 & 15.43 & 0.68 & 5.08 & -50.9 \\
 & 6.76 & 0.64 & 6.64 & -51.29 \\
 & 5.33 & 0.93 & 7.94 & -49.99 \\

\hline

\end{tabular}
\begin{flushleft}
\textbf{Notes.} See Table \ref{tab:table_modelfit}
\end{flushleft}
\end{table}

\begin{table}[]
\centering
\caption{Modelfit parameters of the Gaussian components at 15\,GHz at the different epochs.} 
\begin{tabular}{c|ccccc}
\hline

Obs date & \begin{tabular}[c]{@{}c@{}}S\\ {[}mJy{]}\end{tabular} & \begin{tabular}[c]{@{}c@{}}d\\ {[}mas{]}\end{tabular} & \begin{tabular}[c]{@{}c@{}}z\\ {[}mas{]}\end{tabular} & \begin{tabular}[c]{@{}c@{}}PA\\ {[}deg{]}\end{tabular} \\ \hline

Jul 2009  & 78.78 & 0.13 & 0.05 & -154.25 \\
 & 261.34 & 0.07 & 0.36 & -56.31 \\
 & 150.49 & 0.08 & 0.67 & -54.21 \\
 & 66.26 & 0.19 & 1.1 & -53.6 \\
 & 42.3 & 0.27 & 1.67 & -52.31 \\
 & 25.88 & 0.26 & 2.27 & -52.25 \\
 & 23.1 & 0.65 & 3.11 & -51.13 \\
 & 21.51 & 0.89 & 4.83 & -51.21 \\
 & 8.36 & 0.65 & 6.51 & -51.6 \\
 & 6.32 & 1.13 & 7.89 & -50.18 \\
 & 8.65 & 3.25 & 13.11 & -51.65 \\

\hline

Jan 2010  & 68.52 & 0.09 & 0.06 & -154.17 \\
 & 244.89 & 0.08 & 0.36 & -56.31 \\
 & 174.58 & 0.12 & 0.67 & -55.37 \\
 & 84.0 & 0.16 & 1.09 & -53.2 \\
 & 34.19 & 0.16 & 1.69 & -53.15 \\
 & 31.83 & 0.35 & 2.27 & -52.76 \\
 & 19.05 & 0.71 & 3.31 & -50.6 \\
 & 16.77 & 0.72 & 5.0 & -51.51 \\
 & 11.88 & 1.77 & 7.09 & -51.19 \\
 & 7.77 & 3.96 & 12.3 & -49.35 \\

\hline

Dec 2012  & 96.56 & 0.07 & 0.04 & -135.27 \\
 & 239.64 & 0.09 & 0.36 & -56.31 \\
 & 95.23 & 0.12 & 0.76 & -52.41 \\
 & 79.65 & 0.2 & 1.23 & -52.77 \\
 & 31.95 & 0.23 & 1.76 & -52.09 \\
 & 48.55 & 0.52 & 2.66 & -51.33 \\
 & 15.68 & 0.67 & 3.94 & -52.29 \\
 & 11.69 & 0.68 & 5.03 & -50.25 \\
 & 14.04 & 1.53 & 7.29 & -50.36 \\

\hline

Aug 2019  & 11.49 & 0.04 & -0.36 & 43.98 \\
 & 114.91 & 0.08 & 0.04 & -95.18 \\
 & 152.95 & 0.08 & 0.36 & -56.31 \\
 & 90.99 & 0.1 & 0.79 & -50.44 \\
 & 50.91 & 0.13 & 1.22 & -48.71 \\
 & 22.59 & 0.27 & 1.77 & -49.65 \\
 & 11.45 & 0.23 & 2.34 & -49.92 \\
 & 13.02 & 0.46 & 3.18 & -48.46 \\
 & 15.28 & 0.56 & 3.82 & -49.79 \\
 & 16.67 & 0.63 & 4.82 & -51.53 \\
 & 5.14 & 0.4 & 5.57 & -50.48 \\
 & 12.82 & 2.24 & 7.9 & -50.6 \\

\hline

Jan 2020  & 0.64 & 0.18 & -3.26 & 44.06 \\
 & 0.57 & 0.06 & -1.75 & 44.84 \\
 & 3.56 & 0.02 & -0.66 & 47.93 \\
 & 85.1 & 0.02 & -0.13 & 58.37 \\
 & 197.49 & 0.1 & 0.2 & -45.0 \\
 & 142.5 & 0.08 & 0.64 & -47.88 \\
 & 68.54 & 0.17 & 1.07 & -46.01 \\
 & 31.97 & 0.24 & 1.61 & -47.38 \\
 & 15.94 & 0.21 & 2.12 & -49.5 \\
 & 7.09 & 0.23 & 2.64 & -48.06 \\
 & 15.72 & 0.52 & 3.39 & -48.58 \\
 & 10.51 & 0.53 & 4.1 & -49.4 \\
 & 15.57 & 0.5 & 4.97 & -50.75 \\
 & 5.95 & 0.37 & 5.62 & -49.84 \\
 & 5.53 & 0.71 & 6.8 & -49.26 \\
 & 6.51 & 0.94 & 8.32 & -50.2 \\
 & 2.32 & 0.71 & 9.73 & -50.99 \\
 & 1.87 & 1.29 & 11.62 & -49.38 \\
 & 1.85 & 1.52 & 14.73 & -48.23 \\
 & 0.88 & 1.41 & 21.91 & -53.75 \\
 & 1.18 & 1.22 & 24.86 & -50.83 \\

\hline

\end{tabular}
\begin{flushleft}
\textbf{Notes.} See Table \ref{tab:table_modelfit}
\end{flushleft}
\end{table}

\begin{table}[]
\centering
\caption{Modelfit parameters of the Gaussian components at 22\,GHz at the different epochs.} 
\begin{tabular}{c|ccccc}
\hline

Obs date & \begin{tabular}[c]{@{}c@{}}S\\ {[}mJy{]}\end{tabular} & \begin{tabular}[c]{@{}c@{}}d\\ {[}mas{]}\end{tabular} & \begin{tabular}[c]{@{}c@{}}z\\ {[}mas{]}\end{tabular} & \begin{tabular}[c]{@{}c@{}}PA\\ {[}deg{]}\end{tabular} \\ \hline

Jan 2020  & 49.97 & 0.15 & -0.18 & 49.79 \\
 & 191.39 & 0.09 & 0.08 & -45.0 \\
 & 87.93 & 0.16 & 0.4 & -52.27 \\
 & 87.27 & 0.17 & 0.78 & -47.84 \\
 & 30.54 & 0.22 & 1.23 & -46.79 \\
 & 21.26 & 0.32 & 1.78 & -49.14 \\
 & 10.21 & 0.39 & 2.5 & -48.67 \\
 & 15.44 & 0.65 & 3.61 & -48.89 \\
 & 11.53 & 0.53 & 4.83 & -51.21 \\
 & 6.75 & 0.5 & 5.61 & -50.47 \\
 & 5.06 & 1.22 & 7.56 & -50.36 \\
 & 2.81 & 0.71 & 9.03 & -50.96 \\

\hline

\end{tabular}
\begin{flushleft}
\textbf{Notes.} See Table \ref{tab:table_modelfit}
\end{flushleft}
\end{table}

\begin{table}[]
\centering
\caption{Modelfit parameters of the Gaussian components at 43\,GHz at the different epochs.} 
\begin{tabular}{c|ccccc}
\hline

Obs date & \begin{tabular}[c]{@{}c@{}}S\\ {[}mJy{]}\end{tabular} & \begin{tabular}[c]{@{}c@{}}d\\ {[}mas{]}\end{tabular} & \begin{tabular}[c]{@{}c@{}}z\\ {[}mas{]}\end{tabular} & \begin{tabular}[c]{@{}c@{}}PA\\ {[}deg{]}\end{tabular} \\ \hline

Jan 2020  & 19.29 & 0.01 & -0.26 & 43.41 \\
 & 243.3 & 0.06 & 0.0 & -41.5 \\
 & 41.49 & 0.18 & 0.47 & -49.58 \\
 & 20.68 & 0.01 & 0.83 & -47.59 \\
 & 5.75 & 0.21 & 1.36 & -47.11 \\
 
\hline

Apr 2021  & 23.89 & 0.16 & -0.19 & 53.68 \\
 & 165.74 & 0.08 & 0.0 & -23.47 \\
 & 29.88 & 0.06 & 0.11 & -79.96 \\
 & 22.18 & 0.14 & 0.27 & -51.36 \\
 & 30.89 & 0.21 & 0.57 & -52.41 \\
 & 27.25 & 0.23 & 0.89 & -49.93 \\
 & 17.84 & 0.33 & 1.28 & -50.76 \\
 & 10.9 & 0.46 & 1.97 & -51.33 \\

\hline

Oct 2021  & 36.26 & 0.04 & -0.24 & 43.48 \\
 & 215.45 & 0.04 & 0.0 & -168.51 \\
 & 59.09 & 0.25 & 0.48 & -49.73 \\
 & 36.68 & 0.22 & 0.96 & -48.55 \\
 & 27.32 & 0.88 & 1.89 & -48.56 \\
 & 10.3 & 0.7 & 3.53 & -49.0 \\
 & 5.87 & 0.8 & 4.94 & -48.42 \\
 & 5.05 & 0.51 & 7.67 & -48.64 \\

\hline

Apr 2022  & 5.09 & 0.15 & -0.36 & 51.94 \\
 & 45.21 & 0.04 & -0.12 & 48.7 \\
 & 177.36 & 0.04 & 0.0 & -103.4 \\
 & 60.52 & 0.08 & 0.15 & -53.56 \\
 & 18.55 & 0.1 & 0.44 & -42.31 \\
 & 28.14 & 0.13 & 0.62 & -52.17 \\
 & 26.49 & 0.11 & 0.91 & -47.95 \\
 & 17.01 & 0.19 & 1.19 & -49.51 \\
 & 5.0 & 0.2 & 1.61 & -47.95 \\
 & 5.75 & 0.38 & 2.15 & -49.45 \\
 & 9.17 & 1.54 & 3.61 & -48.89 \\
 & 13.47 & 1.51 & 4.95 & -47.87 \\

\hline

\end{tabular}
\begin{flushleft}
\textbf{Notes.} See Table \ref{tab:table_modelfit}
\end{flushleft}
\end{table}

\end{appendix}

\end{document}